\documentclass[showkeys,preprintnumbers,amsmath,amssymb,nofootinbib]{revtex4}

\usepackage{graphics}
\usepackage{graphicx}
\usepackage{dcolumn}
\usepackage{bm}
\usepackage{epsfig}
\usepackage{wrapfig}
\usepackage{amssymb}
\usepackage{amsmath}

\begin{document}

\title{Relativistic gravitational deflection of light and its impact on the modeling accuracy for the Space Interferometry Mission}

\author{Slava G. Turyshev}

\affiliation{%
Jet Propulsion Laboratory, California Institute of Technology,\\
4800 Oak Grove Drive, Pasadena, CA 91109-0899, USA
}%

\affiliation{%
Sternberg Astronomical Institute, 13 Universitetskij Prospect, 119992 Moscow, Russia\footnote{Email: {\tt turyshev@jpl.nasa.gov, turyshev@sai.msu.ru }}
}%

\date{\today}

\begin{abstract}
We study the impact of relativistic gravitational deflection of light on the accuracy of future Space Interferometry Mission (SIM). We estimate the deflection angles caused by the monopole, quadrupole and octupole components of gravitational fields for a number of celestial bodies in the solar system. We observe that, in many cases, the magnitude of the corresponding effects is significantly larger than the $1~\mu$as accuracy expected from SIM. This fact argues for the development of a relativistic observational model for the mission that would account for the influence of both static and time-varying effects of gravity on light propagation. Results presented here are different from the ones obtained elsewhere by the fact that we specifically account for the differential nature of the future SIM astrometric measurements. We also obtain an estimate for the accuracy of possible determination of the Eddington's parameter $\gamma$ via SIM global astrometric campaign; we conclude that accuracy of $\sim7 \times 10^{-6}$ is achievable via measurements of deflection of light by solar gravity. 
\end{abstract}

\keywords{Interferometric astrometry; SIM; tests of general relativity; solar system.}

\maketitle

\section{Introduction}

The last quarter of the 20th century has changed the status of Einstein's general theory of relativity from a purely theoretical discipline to a practically important science. Today general relativity is the standard theory of gravity, especially where the needs of astronomy, astrophysics, cosmology and fundamental physics are concerned \cite{Turyshev-etal-2007,Turyshev-2008,Soffel-etal-2003}. As such, this theory is used for many practical purposes involving spacecraft navigation, geodesy and time transfer. Present accuracy of astronomical observations already requires relativistic description of light  propagation as well as the relativistically  correct treatment of the dynamics of the extended celestial bodies \cite{Kopeikin-Makarov-2007}. As a result, some of the leading static-field post-Newtonian  perturbations in the dynamics of the  planets, the Moon and artificial satellites have been included in the equations of motion, and in time and position transformation. It is also well understood that effects due to  non-stationary behavior of the solar system gravitational field as well as its  deviation from spherical symmetry should be also considered \cite{Kopeikin-1997} and implemented in the appropriate models.   

Space-based astrometry has brought about a renaissance in the entire field of astrometry that is perhaps the most fundamental, and oldest of all areas in astronomy \cite{Unwin-etal-2008}. The ESA Hipparcos mission, which operated from 1989-1993, yielded an astrometric catalog of 118,000 stars down to 12.5 magnitude, with positional accuracy of 1 mas for stars brighter than V = 11. The European Space Agency (ESA) is now developing the Gaia mission as a next generation astrometric survey mission \cite{Perryman-2001,Perryman-2002}, which is expected to produce a catalog of $\sim 10^9$ stars, with accuracy $\simeq$~20--25 microarcsec ($\mu$as) for stars brighter than V = 15. Precision astrometry remains a cornerstone of the field and is poised to make a major impact on many fields of modern astronomy, astrophysics, and cosmology \cite{Unwin-etal-2008}. 

NASA's SIM PlanetQuest mission, hereinafter SIM, as another example of a space-based facility instrument for astrometry. The acronym SIM stands for Space Interferometry Mission. SIM will be the first space-based Michelson interferometer for astrometry. The instrument will operate in the optical waveband using a 9-m baseline between the apertures. With a global astrometry accuracy of $3~\mu$as for stars brighter than V = 20, it will measure parallaxes and proper motions of stars throughout the Galaxy with unprecedented accuracy. Operating in a narrow-angle mode, it will achieve a positional accuracy
of $0.6~\mu$as for a single measurement, equivalent to a differential positional accuracy at the end of the nominal 5-year mission of = $0.1~\mu$as. This performance is about 1000 times better than existing capabilities on the ground or in space, and about 100 times better than the upcoming Gaia mission, for differential measurements. Such high accuracy will allow SIM to detect and measure masses of terrestrial planets around stars in our Galactic neighborhood (see \cite{Unwin-etal-2008} for review).

SIM is a targeted mission which measures the astrometric positions of stars, referencing the measurements to a grid of 1302 stars covering the entire sky. Its scheduling is highly flexible, in both the order of observations, their cadence, and the accuracy of each individual measurement. This contrasts with the Hipparcos and Gaia missions, which scan the entire sky according to a pre-determined scanning pattern. Many astrometry experiments can make effective use, or in some cases require, this pointing capability – for instance, searches for terrestrial planets (especially in multiple planet systems), stellar microlensing events, orbits of eccentric binary systems, and variable targets such as X-ray binaries and active galactic nuclei. Currently, the ICRF, defined by the locations of 212 extragalactic radio sources \cite{Johnston-etal-1995,Ma-1998} with most having errors less than 1 mas, is the standard frame for astrometry. SIM is expected to yield an optical reference frame at a level of about $3~\mu$as; it will be 'tied' to the ICRF by observing a number of radio-loud quasars in common.

\begin{wrapfigure}{R}{0.58\textwidth}
 \begin{center}
\vskip -15pt
    \includegraphics[width=0.58\textwidth]{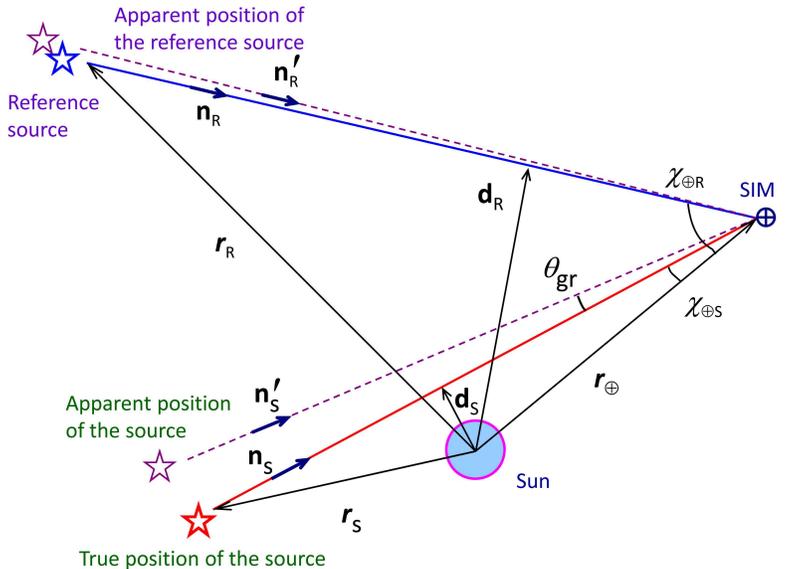}
\vskip -0pt
    \caption{Geometry of gravitational deflection of starlight by the 
Sun. 
      \label{fig:deflect}}
 \end{center}
\vskip -20pt
\end{wrapfigure}

In this paper, we discuss the rapidly forming field of applied general relativity to demonstrate its influence on the high-precision astrometry. 
Recent advances in the accuracy of astrometric observations have demonstrated  importance of taking into account the relativistic effects introduced by the solar system's gravitational environment. It is known that the reduction of the Hipparcos data has necessitated the  inclusion of stellar aberration up to the terms of the second order in $v/c$, and the general relativistic treatment of light bending due to the gravitational field of the Sun \cite{Perryman92} and Earth \cite{Gould93}. Even higher modeling accuracy is anticipated for Gaia \cite{Klioner-2004}.

Prediction of the gravitational deflection of light was one of the first successes of general relativity.  Since the first confirmation by the Eddington's expedition in 1919,  the effect of gravitational  deflection has been studied quite extensively and currently analysis of almost every precise  astronomical measurement must take this effect  into account \cite{Modest96,Brumberg-1972,Brumberg-1991,Turyshev-etal-2007,Turyshev-2008}.
According to general relativity, the  light rays propagating  near a gravitating body are achromatically deflected by the body's relativistic gravity field.  The whole  trajectory of the light ray is bent towards the body by an angle depending on the strength of the body's gravity.   The solar gravity field produces the largest effect on  the light  traversing the solar system. To first order in the gravitational constant, $G$, the  solar deflection  angle $\theta^\odot_{\tt gr}$ depends  only on the solar  mass $M_\odot$ and the  impact parameter $d$ relative to the Sun \cite{Will_book93}:    
{}
\begin{equation}
\theta^\odot_{\tt gr}={4GM_\odot\over c^2
d}\cdot\frac{1+\cos\chi}{2} =1''.751 \Big[\frac{R_\odot}{d}\Big]\frac{1+\cos\chi}{2},
\label{eq:deflec0}
\end{equation}
\noindent 
where $R_\odot$ solar radus. The absolute magnitude for the light deflection angle is maximal for the
rays grazing the sun, e.g. $\theta^\odot_{\tt
gr}= 1.751$ seconds of arc.  Most of the measurements  of the gravitational deflection to date involved the solar gravity field, planets in the solar system  or gravitational lenses. For the future astrometric observations with SIM, in addition to the Sun, effect of planetary gravitatiobal deflections of light must also be considered.  The most precise measurement of the light deflection with the planet Jupiter was done in \cite{Fomalont-Kopeikin03}. Relativistic deflection  of light  has been observed, with various degrees of precision,  on distance scales of $10^9$  to $10^{21}$ m, and on mass scales from  $10^{-3}$  to $10^{13}$ solar masses, the upper ranges determined from the gravitational lensing of quasars \cite{Dar92,TreuhaftLowe91}. 

In the case of SIM, the star is assumed to be at a very large distance compared to the Sun, and $\chi$ is the angular separation between the deflector and the star. With the space observations  carried out by SIM, $\chi$ is not necessarily a small angle.   The relevant geometry and notations are shown in Fig. \ref{fig:deflect}. In this figure we emphasized the fact that the difference of the apparent position of the source from it's true position depends on the impact parameter of the incoming light with respect to the deflector.  For the astrometric accuracy of a few $\mu$as and, in the case when the Sun is the deflector, positions of all observed sources experiencing such a displacement. This is why, in order to correctly account for the effect of gravitational deflection, it is important to process together the data taken with the different separation angles from the deflector. In the wide-angle astrometry mode SIM, will be observing the sky in a 15$^\circ$ patches of sky (called field of regard or FoR) making a set of differential observations within the FoR. Therefore, this differential nature of the measurements would result in minimizing the contribution of the gravitational deflection on the single measurement. To reflect this fact, we will present results for two types of astrometric measurements, namely for the  absolute (single ray deflection) and  differential  (two sources separated by the 15$^\circ$ field of regard) observations. 

A major objective of this paper is to show that, before microarcsecond-level astrometry will become a powerful tool for 21-st century astronomy, there is a need for an adequate modeling necessary to match this new frontier of astrometric accuracy. The prospect of new high precision astrometric measurements from space  with SIM requires inclusion of relativistic effects at the $(v/c)^3$ level \cite{Turyshev98,Unwin-etal-2008}. At the level of accuracy expected from SIM, even more subtle gravitational effects on astrometry from within the solar system will  start to become apparent, such as the monopole and the quadrupole components of  the gravitational fields of the  planets \cite{Sovers98} and the gravito-magnetic effects caused by their  motions and rotations. Thus,  the identification of all possible sources  of ``astrophysical'' noise that may contribute to the future  SIM astrometric campaign, is well justified. 
  
This work is organized as follows: Section
\ref{sec:mon_defl} discusses the influence of the relativistic
deflection of light by the monopole components of the gravitational fields
of the solar system's bodies.  We present the model and our estimates for the most important  effects  that will be influencing 
astrometric observations  of a few $\mu$as accuracy, that will be made from within the solar system.
Section \ref{sec:three_env} will specifically address three 
most intense gravitational environments in the solar system,
namely the vicinities of the Sun, Jupiter and Earth.
In Section \ref{sec:high_multipoles} we will discuss the effects of the
gravitational deflection of light by the higher gravitational multipoles (both mass and current ones) of some of the bodies in the solar system.  
We derive constraints on the navigation of the spacecraft and the accuracy of the solar system ephemerides. 
In Section \ref{sec:astrophys} we investigate the possibility of  improving the accuracy of the Eddington's  parameter $\gamma$ via astrometric tests of general relativity  in the solar system. We also discuss the opportunity to measure the solar acceleration towards the Galactic Center with SIM.
We will conclude the paper with the discussion of the results obtained and  our recommendations for future studies.

\section{Gravity Contributions to the  Local Astrometric Environment}
\label{sec:mon_defl}

In this Section we develop a model for light propagation that will be used to estimate various relativistic effects due to gravitational deflection of light by the solar system's bodies.

\subsection{Relativistic deflection of light by the gravity monopole}

The first step into a relativistic modeling of a light path consists of determining the direction of the incoming photon as measured by an observer located in the solar system as a function of the barycentric coordinate position of the light source. Apart from second and third orders of velocity aberration the only other sizable effect is due to the bending of light rays in the gravitational field of solar system bodies \cite{Brumberg-Klioner-Kopeikin-1990,Turyshev98}. Effects of the gravitational monopole  deflection of light are the largest among those in the solar system.

Generalizing on a phenomenological parameterization of the gravitational metric tensor field, which Eddington originally developed for a special case, a method called the parameterized post-Newtonian (PPN) formalism has been developed (see \cite{Turyshev-2008} for discussion). This method represents the gravity tensor's potentials for slowly moving bodies and weak inter-body gravity, and is valid for a broad class of metric theories, including general relativity as a unique case. The several parameters in the PPN metric expansion vary from theory to theory, and they are individually associated with various symmetries and invariance properties of the underlying theory  (see \cite{Will_book93} for details).

If (for the sake of simplicity) one assumes that Lorentz invariance, local position invariance and total momentum conservation hold, the metric 
tensor for a single, slowly-rotating gravitational source is given by:
{}
\begin{eqnarray}
\label{eq:metric}
g_{00}&=&1-2\frac{GM}{c^2r}\Big(1-J_2\frac{R^2}{r^2}\frac{3\cos^2\theta-1}{2}\Big)+
{\cal O}(c^{-4}),\nonumber\\
g_{0i}&=& 2(\gamma+1)\frac{G[\vec{\cal S}\times \vec{r}]_i}{c^3r^3}+
{\cal O}(c^{-5}),\\\nonumber
g_{ij}&=&-\delta_{ij}\Big[1+2\gamma \frac{GM}{c^2r} \Big(1-J_2\frac{R^2}{r^2}\frac{3\cos^2\theta-1}{2}\Big)+
{\cal O}(c^{-5}),   
\end{eqnarray}
\noindent where $M$ and $\vec{\cal S}$ being the mass and angular momentum of the body, $J_2$ and $R$ are the body's quadrupole moment and its radius, and $G$ is the universal gravitational constant,  $r$ is the distance from the center of the body to a particular point. The $1/c^2$ term in $g_{00}$ is the Newtonian limit; the $1/c^3$ term in $g_{0i}$ and the $1/c^2$ term in $g_{ij}$, are post-Newtonian corrections. All of these terms are required to describe light propagation phenomena to the first post-Newtonian order.  

The Eddington parameter $\gamma$ in the Eqs.~(\ref{eq:metric}) represents the measure of the curvature of the space created by a unit rest mass \cite{Will_book93}. Note that general relativity, when analyzed in  standard gauge of the PPN formalism (see \cite{Will_book93,Will-lrr-2006-3} for details), gives: $\gamma=1$. The Brans-Dicke theory is  the most famous among the alternative theories of gravity.  It contains, besides the metric tensor,  a scalar field $\phi$ and an arbitrary coupling constant $\omega$, related to this PPN parameter as $\gamma= {1+\omega\over 2+\omega}$.  The stringent observational bound resulting from the 2003 experiment with the Cassini spacecraft require that $|\omega| \gtrsim 40000$ \cite{Bertotti-Iess-Tortora-2003,Will-lrr-2006-3}. There exist additional alternative theories that provide guidance for gravitational experiments \cite{Will-lrr-2006-3}.

Metric tensor Eq.~(\ref{eq:metric}) can be used to derive expressions need to describe propagation of electro-magnetic signals between any of the two points in space.  Following the standard procedure of integrating light geodesics (see \cite{Brumberg-1972,Brumberg-1991} for details), the corresponding light-time equation for a single deflecting body can be derived in the following form 
\begin{equation}
\label{eq:light-time}
t_2-t_1= \frac{r_{12}}{c}+(1+\gamma) \frac{G M}{c^3}
\ln\left[\frac{r_{1}+r_{2}+r_{12}}{r_{1}+r_{2}-r_{12}}\right]+{\cal O}(c^{-5}),
\end{equation}

\noindent  where $t_1$ refers to the signal transmission time, and $t_2$ refers to the reception time.  $r_{1,2}$ are the barycentric positions of the transmitter and receiver, and $r_{12}$ is their spatial separation (see \cite{Moyer-2003} for details). Also, for the moment, we neglected the presence of $J_2$ and $\vec{\cal S}$ in the Eqs.~(\ref{eq:metric}), but will investigate contributions of these and other multipoles in Sec.~\ref{sec:high_multipoles}.

For a realistic observing scenario with the SIM, the sources of light are located far out side the solar system, $r_2\ll r_1\equiv r_S$ and $r_{12}$ can be approximated as $r_{12}=|\vec{r}_1-\vec{r}_2|\simeq r_1-(\vec{n}_S\cdot \vec{r}_2)$, where we introduced a notation $\vec{n}_S=\vec{r}_1/r_1$. This approximation allows one to represent the expression in the square brackets in Eq.~(\ref{eq:light-time}) as follows:
\begin{equation}
\label{eq:light-time-approx}
\frac{r_1+r_2+r_{12}}{r_1+r_2-r_{12}}\simeq 
\frac{2r_1+r_2-(\vec{n}_S\cdot \vec{r}_2)}{r_2+(\vec{n}_S\cdot\vec{r}_2)}.
\end{equation}

\begin{wrapfigure}{R}{0.5\textwidth}
 \begin{center}
\vskip -20pt
    \includegraphics[width=0.5\textwidth]{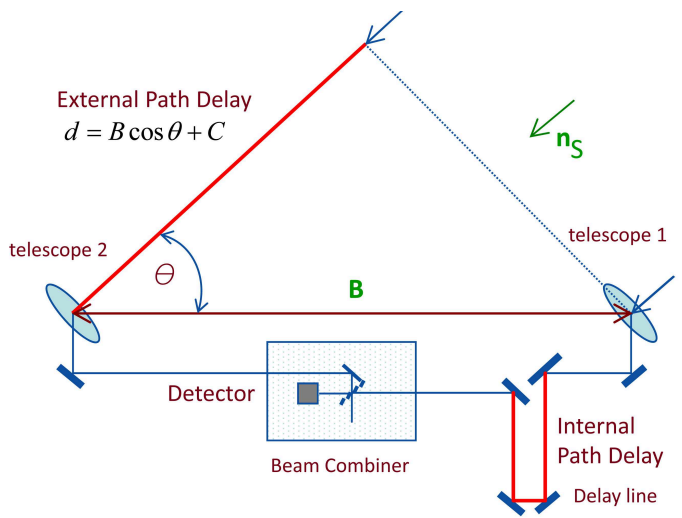}
\vskip -10pt
    \caption{Basic geometry of light propagation in the stellar interferometry. 
      \label{fig:astrom}}
 \end{center}
\vskip -10pt
\end{wrapfigure}

The SIM instrument does not directly measure the angular separation between stars, but the projection of each star direction vector onto the interferometer baseline by measuring the pathlength delay of starlight as it passes through the two arms of the interferometer. The SIM instrument will precisely measure optical path difference (OPD) between the wavefronts of light received by the two telescopes forming the interferometric baseline (see Fig.~\ref{fig:astrom}). (In fact, for SIM, with its 10 m baseline, such a measurement will be done with a precision at the level of 1 picometer.) This difference will result in the different internal OPDs needed to apply in order to coherently add the signals. The delay measurement is made by a combination of internal metrology measurements to determine the distance the starlight travels through each arm,  external metrology measurements that determine the length and local orientation of the baseline, and a measurement of the central white light fringe to determine the point of equal optical pathlength \cite{Unwin-etal-2008}.  Therefore, the OPD is the main observable that the interferometric instrument will measure; relativistic modeling this delay will be among the main objectives of the upcoming SIM modeling effort. 

An interferometer measures optical path difference (OPD) between the wavefronts of light received by the two telescopes forming the interferometric baseline $\vec{b}$ at points $\vec{r}_2$ and $\vec{r}'_2=\vec{r}_2+\vec{b}$.  To account for that fact we need to determine the temporal difference between the signals received at these telescopes which is  $\ell=c(t'_2-t_1)-c(t_2-t_1)=c(t'_2-t_2)$. The first term in Eq.~(\ref{eq:light-time}) is the geometric delay. Using approximations $(b\ll r_2\ll r_1)$, it is easy to see that this term leads to the approximate expression for geometric delay, given as  $\ell_{\rm geom}=r_{12}'-r_{12}=-(\vec{b}\cdot \vec{n}_S)/(1-\vec{n}_S\cdot \dot{\vec{r}}_2/c)$. In this paper, we concern with only the largest contributions from the gravitational defection of light, thus most of the velocity-dependent terms will discarded (see \cite{Sovers98,Kopeikin-Schafer-1999} for details).

The second term Eq.~(\ref{eq:light-time}) is the relativistic delay $\ell_{\rm gr}$, the focus of this work. Using Eq.~(\ref{eq:light-time-approx}) one can present expression in the square brackets of Eq.~(\ref{eq:light-time}) as below 
{}
\begin{equation}
\label{eq:delay}
\frac{2r_1+{r_2}'-(\vec{n}_S\cdot \vec{r}_2{}')}{2r_1+r_2-(\vec{n}_S\cdot \vec{r}_2)}~\frac{r_2+(\vec{n}_S\cdot\vec{r}_2)}{{r_2}{}'+(\vec{n}_S\cdot\vec{r}_2{}')}\simeq
1-\frac{1}{r_2}\,\frac{\vec{b}(\vec{n}_S+\vec{n}_2)}{1+(\vec{n}_S\cdot\vec{n}_2)}.
\end{equation}

In the first order in gravitational constant, one can add individual interferometric delays due to the gravity of the bodies along the light path. As a result, the general relativistic contribution to the OPD $\ell_{\tt gr}=c\tau_{\tt gr}$ takes the following approximate form: 
{}
\begin{equation}
\ell_{\tt gr}= -(\gamma+1)\sum_B\frac{G}{c^2}\frac{M_B}{r_B}
\Big[{\vec{b}(\vec{n}_S+\vec{n}_B)\over 
1+(\vec{n}_S\cdot\vec{n}_B)}\Big], 
\label{eq:defl0}
\end{equation}
\noindent where $r_B$ is the distance from SIM to a deflecting body $B$, $\vec{n}_B={\vec r}_B/r_B$  is the unit vector in this direction.  This OPD is the leading general relativistic observable that the interferometric instrument will measure; a complete relativistic modeling this delay should be the main objective of the upcoming SIM modeling effort. 

\begin{wrapfigure}{R}{0.6\textwidth}
 \begin{center}
\vskip -10pt
    \includegraphics[width=0.6\textwidth]{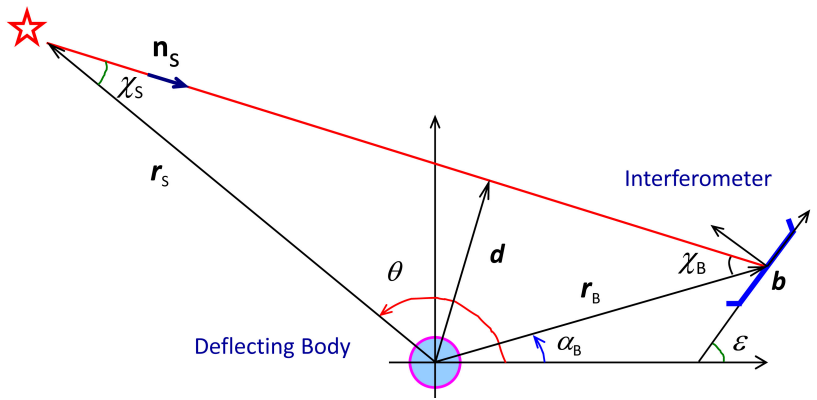}
\vskip -0pt
    \caption{Geometry and notations for the 
gravitational deflection of light. 
      \label{fig:defl_param}}
 \end{center}
\vskip -10pt
\end{wrapfigure}

In general, a three-dimensional approach must be used in order to work out a practical model of the interferometric time delay. Howevere, for the purpose of this paper, it is sufficient to confine our analysis 
to a plane and parameterize  the quantities involved  as
follows (see Fig.~\ref{fig:defl_param}): 
\begin{equation}
\vec{b}= b \,(\cos\epsilon,\, \sin\epsilon), ~~~~~
\vec{r}_B= r_B\,(\cos\alpha_B, \,\sin\alpha_B), ~~~~~
\vec{n}_S= (\cos\theta,\, \sin\theta),
\label{eq:angles}
\end{equation}

\noindent where $\epsilon$ is the angle of the baseline's
orientation with respect to the instantaneous body-centric coordinate
frame,  $\alpha_B$ is the right assention angle of the interferometer
as seen from the this frame and $\theta$ is the direction to the
observed source correspondingly. The geometry of the problem and notations are presented in the Figure \ref{fig:defl_param}. 

It is convenient to express the gravitational contribution to the
total OPD Eq.(\ref{eq:defl0}) in terms of the deflector and the source separation angle $\chi_B$ as observed by the interferometer. In our approximation, the following relations $d=r_S\sin\chi_S=r_B\sin\chi_B$ and $\chi_S+\chi_B+\theta-\alpha=\pi$ are valid; this allows one to eliminate angle $\chi_S$ by expressing the source's position angle $\theta$ via the  separation angle $\chi_B$ as below:
\begin{equation}
\theta=\pi+\alpha_B-\chi_B-\arcsin\Big[\,\frac{r_B}{r_S}\,
\sin\chi_B\,\Big].
\label{eq:theta_chi}
\end{equation}
\noindent As the sources will be located at a very large distance, $r_S$, compares to the distance between the interferometer and the deflector ($r_B\ll r_S$), we can neglect the presence of the last term in the equation Eq.(\ref{eq:theta_chi}), so that 
$\theta\simeq\pi+\alpha_B-\chi_B$.
After substituting expressions (\ref{eq:theta_chi}) and (\ref{eq:angles}) into Eq.~(\ref{eq:defl0}), we  rewrite the contribution of the gravitational deflection to the total OPD, Eq.(\ref{eq:defl0}),  in the following form: 
{}
\begin{equation}
\ell_{\tt gr}= -(\gamma+1)\sum_B\frac{G}{c^2}\frac{M_Bb}{r_B}
\Big[\cos(\epsilon-\alpha_B)+\sin(\epsilon-\alpha_B)
\frac{1+\cos\chi_B}{\sin\chi_B}\Big].
\label{eq:defl_chi}
\end{equation}
This expression describes gravitational delay as measured by an interferometer; we will use it for estimation purposes. 

\subsubsection{Absolute Astrometric Measurements}

Eq.~(\ref{eq:defl_chi}) is appropriate for estimation the magnitudes of the gravitational bending effects measured by an interferometer. It depends on the angle between the baseline and deflector-instrument vectors, $\epsilon-\alpha_B$. As the main objective of this paper is to estimate the magnitudes of the effects involved, we choose $\epsilon-\alpha_B=\frac{\pi}{2}$ that maximizes  contribution of each individual deflector for a particular orbital position of the spacecraft and the baseline orientation.  
As a result, in the SIM proper reference frame Eq.~(\ref{eq:defl_chi}) may be re-written as $\ell_{\tt gr}=-\sum_B \,\ell_{\tt gr}^B$, with the individual contributions of the deflecting bodies to gravitational delay $\ell_{\tt gr}^B$ and  deflection angle  $\theta^B_{\tt gr}\simeq{\ell^B_{\tt gr}}/{b}$ in the following form
{}
\begin{equation}
\ell_{\tt gr}^B= -(\gamma+1)\frac{G}{c^2}\frac{M_Bb}{r_B}
\frac{1+\cos\chi_B}{\sin\chi_B} \qquad {\rm and} \qquad 
\theta^B_{\tt gr}= -(\gamma+1)\frac{G}{c^2}\frac{M_B}{r_B}
\frac{1+\cos\chi_B}{\sin\chi_B}. 
\label{eq:defl_chi-ind}
\end{equation}
\noindent
The two expressions $\ell_{\tt gr}^B$ and $\theta^B_{\tt gr}$ will be used interchangeably throughout the paper. 

For complete analysis of the gravitational deflection of light we will have to account for the time dependency in all the quantities involved. Thus, one will have to use the knowledge of the position of the spacecraft in the solar system's barycentric reference frame, the instrument's orientation in the proper coordinate frame \cite{Klioner03}, the time that was spent in a particular orientation, the history of all the maneuvers and re-pointings of the instrument, etc. These issues are closely related to the principles of the operational mode of the instrument that is currently still being developed.  

\subsubsection{Differential Astrometric Measurements}

SIM will perform its astrometric campaign working in differential mode either within FoR=15$^\circ$ for the wide angle astrometry or within FoR=1$^\circ$ for the narrow angle astrometry.   To evaluate the impact of gravitational delay of light on these measurements, we need to derive the appropriate expressions reflecting the differential nature of astrometric measurements with SIM.

Within the accepted approximation, the necessary expression for the differential OPD may be obtained by subtracting OPDs for the different sources one from one another. Using Eq.~(\ref{eq:defl_chi}), this results in the following expression:   
{} 
\begin{equation}
\delta\ell^B_{\tt gr} = \ell^B_{1\tt gr}-\ell^B_{2\tt gr}=
-(\gamma+1) \sum_B\frac{G}{c^2}\frac{M_B}{r_B}
\Big[\frac{\vec{b}(\vec{n}_{S1}+\vec{n}_B)}
{1+(\vec{n}_{S1}\vec{n}_B)}-
\frac{\vec{b}(\vec{n}_{S2}+\vec{n}_B)}
{1+(\vec{n}_{S2}\vec{n}_B)}\Big], 
\label{eq:delay}
\end{equation}
 
\noindent where ${\vec n}_{S1}$ and $\vec{n}_{S2}$ are the  barycentric positions of the primary and the secondary objects. By using parameterization for the  quantities involved similar to that above ($b\ll r_B \ll r_{S1,S2} $), this expression may be presented in terms of the  deflector-source separation angles, $\chi_{1B}, ~\chi_{2B}$,  as follows:
\begin{equation}
\delta\ell_{\tt gr}= -(\gamma+1)\sum_B\frac{G}{c^2}\frac{M_Bb}{r_B}
\sin(\epsilon-\alpha_B)\,
 \,\frac{\sin\frac{1}{2}(\chi_{B2}-\chi_{B1})}
{\sin\frac{1}{2}\chi_{B1}\sin\frac{1}{2}\chi_{B2}}. 
\label{eq:defl_chi_diff}
\end{equation}

\noindent Similar to the discussion of absolute defection angles, we choose $\epsilon-\alpha_B=\frac{\pi}{2}$ that maximizes  contribution of each individual deflector for a particular orbital position of the spacecraft and the baseline orientation. Therefore, in the SIM proper reference frame Eq.~(\ref{eq:defl_chi_diff}) may be re-written as $\delta\ell_{\tt gr}=-\sum_B \,\delta\ell_{\tt gr}^B$, with the individual contributions for gravitational delay $\delta\ell_{\tt gr}^B$  and corresponding deflection angle $\delta\theta^B_{\tt gr}\simeq\delta{\ell^B_{\tt gr}}/{b}$ in the following form
\begin{equation}
\delta\ell^B_{\tt gr}=
(\gamma+1)~\frac{G}{c^2}\frac{M_B b}{r_B}~
\frac{\sin\frac{1}{2}(\chi_{2B}-\chi_{1B})}
{\sin\frac{1}{2}\chi_{1B}\cdot\sin\frac{1}{2}\chi_{2B}} \qquad {\rm and} \qquad
\delta\theta^B_{\tt gr}=
(\gamma+1)~\frac{G}{c^2}\frac{M_B}{r_B}~
\frac{\sin\frac{1}{2}(\chi_{2B}-\chi_{1B})}
{\sin\frac{1}{2}\chi_{1B}\cdot\sin\frac{1}{2}\chi_{2B}}.
\label{eq:diff_opd}
\end{equation}

\noindent The two expressions $\delta\ell^B_{\tt gr}$ and $\delta\theta^B_{\tt gr}$ will be used interchangeably throughout the paper. 

\subsection{Deflection of Grazing Rays by the Bodies of the Solar
System}

\begin{table} 
\begin{center}
\caption{Relativistic monopole deflection of by the solar system bodies at the SIM's location.
} 
\label{tab:mon}
\vskip 10pt 
\begin{tabular}{|r|r|r|r|r|} \hline

Solar  & Angular size   & 
\multicolumn{3}{c|}{Deflection of grazing rays}
\\[2pt]\cline{3-5}       
system's & at SIM pos., &  absolute  &
diff. $[15^\circ]$ & diff. $[1^\circ]$ \\[2pt]
object & ${\cal R}_B$, arcsec &  $\theta^B_{\tt gr}, \mu$as  &
$\delta\theta^B_{\tt gr},~\mu$as & 
$\delta\theta^B_{\tt gr}, ~\mu$as \\[2pt]\hline\hline

Sun      & 0$^\circ$.26656 & 1$''$.75064   
         & 1$''$.72025 & 1$''$.38221 \\

Sun at 45$^\circ$& 45$^\circ$ & 9\,831.39  
         & 2\,777.97 &  237.66 \\

Moon     & 47.92690 & 25.91 &  25.87 & 25.56 \\

Mercury  & 5.48682  & 82.93  & 82.92  & 82.81 \\

Venus    & 30.15040 & 492.97  & 492.69 &  488.88 \\

Earth    & 175.88401 & 573.75  & 571.90 & 547.03 \\
 
Mars     & 8.93571 & 115.85 & 115.83 & 115.57 \\

Jupiter  & 23.23850 & 16\,419.61   & 16\,412.60 & 16\,314.30 \\

Jupiter at 30$''$ & 30.0 & 12\,719.12   & 12\,712.03 & 12\,614.21 \\

Saturn   & 9.64159 & 5\,805.31 & 5\,804.27  & 5\,789.79 \\

Uranus   & 1.86211 & 2\,171.38  & 2\,171.30 & 2\,170.26 \\

Neptune  & 1.18527 & 2\,500.35  & 2\,500.29  & 2\,499.52 \\
 
Pluto    & 0.11478 & 2.82  & 2.82 & 2.82 \\\hline

\end{tabular} 
\end{center} 
\end{table}

We are now ready to evaluate  the influence of the solar system's gravity field on the future high-accuracy astrometric observations.  In particular, we estimate the magnitudes of the angles of gravitational deflection for those light rays that are grazing the surfaces of celestial bodies.

Table \ref{tab:mon} shows magnitude of the angles characterizing relativistic monopole deflection of grazing (e.g. $\chi_{1B}={\cal R}_B$) light rays by the bodies of the solar system at the SIM's location (i.e., the solar Earth-trailing orbit \cite{Unwin-etal-2008}). Results for absolute deflection angles agree with values obtained by other authors (for instance, \cite{Brumberg-Klioner-Kopeikin-1990}). The results presented in the terms of the following quantities: 
\begin{itemize}
\item[i).] for absolute astrometry results are given 
in terms of the absolute measurements  
$\ell^B_{\tt gr}$ and $\theta^B_{\tt gr}$ from Eq.~(\ref{eq:defl_chi-ind});
\item[ii).] for differential astrometry results are given 
in terms of the absolute measurements  
$~\delta\ell^B_{\tt gr}$ and $\delta\theta^B_{\tt gr}$ from Eq.~(\ref{eq:diff_opd}). 
\end{itemize}

For the differential observations the two stars are assumed to be separated by the size of the instrument's field of regard. For the grazing rays, position of the primary star is assumed to be on the limb of the deflector. Moreover, results are given for the smallest distances from SIM to the bodies (e.g.when the gravitational deflection effect is largest). For the Earth-Moon system we took the SIM's position at the end of the first half of the first year mission at the distance of 0.05 AU from the Earth.  Presented in the right column of Table~\ref{tab:mon} are the magnitudes of the body's individual contributions to the  gravitational  delay of  light  at the SIM's location.

Note, that the angular separation of the
secondary star will always be taken larger than that for the primary. It is convenient to study the case of the most distant available separations of the sources. In the case of SIM, this is the  size of the  field of regard (FoR). Thus for the wide-angle astrometry the size of  FoR will be $15^\circ\equiv\frac{\pi}{12}$~rad, thus $\chi_{2B}=\chi_{1B}+\frac{\pi}{12}$. For the narrow-angle observations this size is FoR $=1^\circ\equiv\frac{\pi}{180}$~rad, thus for this type of astrometric observations we will use $\chi_{2B}=\chi_{1B}+\frac{\pi}{180}$; finally, $b=10$ m is the baseline
length used in the estimates.

\subsubsection{Critical Impact Parameter for High Accuracy Astrometry}
 
The estimates, presented in the Table \ref{tab:mon} have demonstrated that it is very important to correctly model and account for gravitational influence of the bodies of the solar system. Depending on the impact parameter $d_B$ (or planet-source separation angle, $\chi_B$), one will have  to account for the   post-Newtonian deflection of  light by a particular planet. Most important is that one will have to permanently monitor the presence of some of the bodies of the solar system during all astrometric observations, independently on the position of the spacecraft in it's solar orbit and the observing direction. The bodies that introduce the biggest astrometric inhomogeneity
are the Sun, Jupiter and the Earth (especially at the beginning of
the mission, when the spacecraft is in the Earth' immediate proximity).

\begin{table} 
\begin{center}
\caption{Relativistic monopole deflection of light: the angles
and the critical distances for $\Delta\theta_0=1~\mu$as astrometric
accuracy. Solar defection of light must be always taken into account, as at the SIM's position at 1 AU from the Sun, the solar gravitational deflection effects are always larger than 1~$\mu$as. The critical distances for the Earth are given for two distances, namely for 0.05 AU (27$^\circ$.49) and 0.01 AU (78$^\circ$.54).
\label{tab:crit_dist}}
\vskip 10pt 
\begin{tabular}{|r|c|r|c|c|} \hline

Object & $\theta^B_{\tt gr}, ~\mu$as &
\multicolumn{3}{c|}{Critical distances 
for  accuracy of ~$1~\mu$as}
\\[2pt]\cline{3-5}   
  &   & $d^B_{\tt crit}$,~km & $d^B_{\tt crit}$, ~deg & 
$d^B_{\tt crit}$, ~${\cal R}_B$\\\hline\hline

Sun     & 1$''$.75064    &   
          always~~~ & always & always\\


Moon    &  25.91  & $4.501 \times 10^4$ & 
           $0^\circ.34 - 1^\circ.72$ & $25.9 \cdot {\cal R}_m$\\

Mercury & 82.93  & $2.023 \times 10^{4}$ & 
          $0^\circ.06 - 0^\circ.13$ &
          $82.9\cdot {\cal R}_{Me}$\\

Venus   & 492.97  & $2.982 \times 10^{6}$ &
          $0^\circ.66 - 4^\circ.13$  & $492.9\cdot {\cal R}_V$\\

Earth   & 573.75 & $3.453 \times 10^{6}$ & 
          27$^\circ.49-78^\circ$.54 &  
          $ 541.4\cdot {\cal R}_\oplus$\\
 
Mars    & 115.85 & $3.931\times 10^{5}$ &
           0$^\circ.06-0^\circ$.29 & $115.9\cdot {\cal R}_{Ms}$\\

Jupiter & 16\,419.61 & $6.270\times 10^{8}$ & 
          $64^\circ.06-88^\circ.51$ & 
          $8\,849 \cdot {\cal R}_J$\\

Saturn  & 5\,805.31  & $3.420 \times 10^{8}$ & 
          $12^\circ.56-15^\circ.45$ & 
          $5\,700 \cdot {\cal R}_S$ \\

Uranus  & 2\,171.38 & $5.319\times 10^{7}$ & $1^\circ.01-1^\circ.12$ & 
          $2\,171\cdot {\cal R}_U$ \\

Neptune  & 2\,500.35 & $6.276 \times 10^{7}$ & 
                          $0^\circ.77-0^\circ.82$  & 
           $2\,500\cdot {\cal R}_N$ \\
 
Pluto    & 2.82  & $9.025\times 10^3$ & $0''.31-0''.32 $  
         &  $2.8 \cdot {\cal R}_P$\\\hline

\end{tabular} 
\end{center} 
\end{table}

Let us introduce a measure of such a gravitational inhomogeneity
due to a particular body in the solar system. To do this,  suppose that future astrometric experiments with SIM  will be capable to measure astrometric parameters with  accuracy of $\Delta \theta_0=\Delta k ~\mu$as, where $\Delta k$ is some number characterizing the accuracy of the instrument (e.g. for a single measurement accuracy $\Delta k =3$ for stars brighter than $V=20$ and for the mission accuracy $\Delta k=0.1$, see \cite{Unwin-etal-2008}). Then, there will be a critical distance from the body, beginning from which, it is important to account for the presence of the body's gravity in the vicinity of the observed part of the sky. We call this distance -- critical impact parameter, $d^B_{\tt crit}$, the closest distance between the body and the light ray that is gravitationally deflected to the angle 
\begin{equation}
\theta^{\tt c}_{\tt gr}(d^B_{\tt crit}) = \Delta\theta_0 = \Delta k
~\mu{\sf as}.
\label{eq:crit_condition}  
\end{equation}

The necessary expression for $d^B_{\tt crit}$ is obtained from Eq.~(\ref{eq:defl_chi-ind}).  Assuming that the angle $\chi_B$ is small and noting that $r_B\sin\chi_B=d$, we can write this equation as follows  $\theta^B_{\tt gr}\simeq {2\mu_B}/{d}$, where $\mu_B=2GM_B/c^{2}$ being relativistic gravitational radius of the body. As the effect of gravitational deflection light is inversely proportional to the impact parameter, then beginning from a certain value of the parameter, $d^B_{\tt crit}$, the deflection angle will be larger $\Delta \theta_0$; this value is given by the following expression:  
\begin{equation}
d^B_{\tt crit} = \frac{2\,\mu_B}{\Delta\theta_0},
\label{eq:crit_dist}
\end{equation}

Different forms of the critical impact parameters
$d^B_{\tt crit}$ for $\Delta\theta_0=1~\mu$as are given in the 
Table \ref{tab:crit_dist}. With the help of  
Eq.~(\ref{eq:crit_dist}), the results given in
this table are easily scaled for any astrometric accuracy 
$\Delta\theta_0$.

\subsubsection{Deflection of Light by Planetary Satellites}

One may expect that the planetary satellites will affect the astrometric studies if a light ray would pass in their vicinities.  Just for completeness of our study we would like to present the estimates for the gravitational deflection of light by the  planetary satellites and the small bodies in the solar system. The corresponding estimates for deflection angles, $\theta^B_{\tt gr}$, and   critical distances, $d_{\tt crit}$ are presented  in the Table \ref{tab8}. Due, to the fact that the angular sizes for those bodies are much less than the smallest field of regard of the SIM instrument (e.g. FoR=1$^\circ$), the results for the differential observations will be effectively insensitive to the size of the the two available FoRs. The obtained results  demonstrate the fact that observations of these objects with that size of FoR will evidently have the effect from the relativistic bending of light. Thus, in Table \ref{tab8} we have presented there only the angle for the absolute gravitational deflection in terms of quantities $\theta^B_{\tt gr}$.  As a result, the major satellites of Jupiter, Saturn and Neptune should also be included in the model   if the light ray  passes close to these bodies. 

\begin{table} 
\begin{center}
\caption{Relativistic deflection of  light by some
 planetary satellites.}\label{tab8} 
\vskip 10pt 
\begin{tabular}{|r|c|c|c|r|r|r|} \hline

Object &  Mass,   & Radius,   & Angular size,  &
Grazing & 
\multicolumn{2}{c|}{1 $\mu$as critical radius}\\\cline{6-7}
&    $10^{25} $~g &  ${\cal R}_B$, ~km &  ${\cal R}_B$, arcsec &
$\theta^B_{\tt gr}, ~\mu$as &  $d_{\tt crit}, $~km &  $d_{\tt crit},$ 
${\cal R}_{planet}$\\\hline\hline

Io    & 7.23 & 1\,738 & 0.570056 & 25.48 & 44\,291
            &  $0.63\cdot{\cal R}_J$ \\ 
 
Europa  & 4.7 & 1\,620 & 0.531353  & 17.77
        & 28\,793 & $0.41\cdot{\cal R}_J$ \\

Ganymede  & 15.5 & 1\,415 & 0.464114 & 67.11 & 94\,954
          & $1.34\cdot{\cal R}_J$ \\  

Callisto  & 9.66 & 2,450 & 0.803589& 24.15 
          &59\,178& $0.84\cdot{\cal R}_J$ \\ 
 
Rhea  & 0.227 & 675 & 0.108468 & 2.06 & 1\,391  
      & $0.02\cdot{\cal R}_S$ \\  

Titan  & 14.1 & 2\,475 & 0.397715 & 34.90  & 86\,378 
       &$1.44\cdot{\cal R}_S$ \\ 
 
Triton  & 13 & 1\,750 & 0.082638 & 45.51 & 79\,639
        &$3.17\cdot{\cal R}_N$ \\ \hline

\end{tabular} 
\end{center} 
\end{table}

\subsubsection{Gravitational Influence of Small Bodies}
 
Additionally,  for the astrometric accuracy at the level of few $\mu$as (i.g., $\Delta\theta_0=\Delta k~\mu{\rm as}$), one needs to  account for the post-Newtonian deflection of  light due to rather a large number of small bodies in the solar system having a mean
radius  {}
\begin{equation}
 {\cal R}_B \ge  624 ~
\sqrt{\Delta k\over \rho_B}\hskip 8pt ~{\rm km}. 
\label{(2.11)}
\end{equation}

The deflection angle for the largest asteroids Ceres, Pallas and Vesta for $\Delta k=1$ are given in the Table \ref{tab7}.
The quoted properties of the asteroids were taken from  
\cite{StandishHellings89,Mouret-Hestroffer-Mignard-2007}. Positions of these asteroids are  known and they  are incorporated in the JPL ephemerides. Other small bodies (e.g. asteroids, Kuiper belt objects, etc.) may produce a stochastic noise in the future astrometric observations with SIM; therefore,  they should also be properly modeled. 
\begin{table}[h]
\begin{center}
\caption{Relativistic deflection of light by the asteroids. 
\label{tab7}} 
\vskip 10pt
\begin{tabular}{|r|c|c|c|} \hline

Object   & $\rho_B,$ ~g/cm$^3$
& Radius,  km & $\theta^B_{\tt gr}, ~\mu$as \\\hline\hline

Ceres       & 2.3& 470 &1.3 \\ 
Pallas      & 3.4 & 269 & 0.6 \\ 
Vesta       & 3.6 & 263 & 0.6 \\ 
Class S     & 2.1 $\pm$ 0.2 & TBD & $\le 0.3$\\ 
Class C     & 1.7 $\pm$ 0.5 & TBD & $\le 0.3$\\\hline 

\end{tabular} 
\end{center} 
\end{table}

\section{Regions with the most gravitationally intense environments for SIM}
\label{sec:three_env}

The properties of the solar system's gravity field presented in the
Tables \ref{tab:mon} and \ref{tab:crit_dist} suggest that the most
intense gravitational environments   in the solar system are those
offered by the Sun and two planets, namely the Earth and Jupiter. In this Section we will analyze these regions in more details. 

\subsection{Gravitational Deflection of Light by the Sun} 

From the  expressions Eq.~(\ref{eq:defl_chi}) and 
Eq.~(\ref{eq:diff_opd}) we obtain the relations for relativistic
deflection of light by the solar gravitational monopole. The expression
for the   absolute astrometry takes the form:   
{} 
\begin{equation}
\theta^\odot_{\tt gr}= (\gamma+1)~\frac{G}{c^2}
\frac{M_\odot}{r_{\rm AU}}~\frac{1+\cos\chi_{1\odot}}{\sin\chi_{1\odot}}=
4.072\cdot\frac{1+\cos\chi_{1\odot}}{\sin\chi_{1\odot}} ~~~~{\rm mas},
\label{eq:sun_defl}
\end{equation}
\noindent where $\chi_{1\odot}$ is the Sun-source
separation angle,  $r_{\rm AU}=1$~AU, and $\gamma=1$.
Similarly, for differential  astrometric observations  one obtains:
\begin{equation}
\delta\theta^\odot_{\tt gr}= (\gamma+1)\frac{G}{c^2} ~
\frac{M_\odot}{r_{\rm AU}}\frac{\sin\frac{1}{2}(\chi_{2\odot}-\chi_{1\odot})}
{\sin\frac{1}{2}\chi_{1\odot}\cdot\sin\frac{1}{2}\chi_{2\odot}} 
=4.072\cdot\frac{\sin\frac{1}{2}(\chi_{2\odot}-\chi_{1\odot})}
{\sin\frac{1}{2}\chi_{1\odot}\cdot\sin\frac{1}{2}\chi_{2\odot}}~~~~
{\rm mas}, 
\label{eq:sun_def_defl}
\end{equation}
\noindent with $\chi_{1\odot}, \chi_{2\odot}$ being the Sun-source separation angles for the primary and the secondary stars correspondingly. We use two stars separated by the SIM's field of regard, namely $\chi_{2\odot}=\chi_{1\odot}+\frac{\pi}{12}$.  The solar angular dimensions from the Earth' orbit are  calculated to be ${\cal R}_\odot=0^\circ.26656$. This angle corresponds to a deflection of light to $1.75065 $~arcsec 
on the limb of the Sun.  Results for the most interesting  range of $\chi_{1\odot}$   are given in the Table \ref{tab:sun}. 

\begin{table}
\begin{center}
\caption{Magnitudes of the gravitational deflection angle vs. the
Sun-source separation angle $\chi_{1\odot}$.}
\label{tab:sun}
\end{center}
\begin{center}
\vskip -10pt
\begin{tabular}{|r|c|c|c|c|c|c|c|}
\hline
\rm Solar    & \multicolumn{7}{c|}{small $\chi_{1\odot},$ ~deg}\\[1pt]
\cline{2-8}
\rm deflection  &$0^\circ.27$ &$0^\circ.5$&$1^\circ$&$2^\circ$&
$5^\circ$&$10^\circ$&$15^\circ$\\\hline  
$~\theta^\odot_{\tt gr}$, ~mas &
1\,728
& 
933.295 & 
466.639 &
233.302 & 
93.271 & 
46.547 & 
30.932 \\[3pt] \hline
$\delta\theta^\odot_{\tt gr}~[15^\circ]$, ~mas &
1\,698 &
903.372& 
437.663 &
206.053 &
70.176 &
28.178 & 
15.734 \\[3pt] \hline
$\delta\theta^\odot_{\tt gr}~[1^\circ]$, ~mas &
1\,361 &
622.212& 
233.337 &
77.787 &
15.567 &
4.254 & 
1.956 \\[3pt] \hline
\end{tabular} 
\vskip 1pt
\begin{tabular}{|r|c|c|c|c|c|c|c|c|}
\hline
\rm Solar    & \multicolumn{8}{c|}{large $\chi_{1\odot},$ ~deg }  
\\[1pt]\cline{2-9} 
\rm deflection & $20^\circ$ & $40^\circ$ & $45^\circ$ & $50^\circ$
& $60^\circ$ & $70^\circ$ & $80^\circ$ & $90^\circ$\\\hline 
$~\theta^\odot_{\tt gr}$, ~mas &
23.095&
11.189 & 
9.832 &
8.733 &
7.053 &
5.816 &
4.853 &
4.072 \\[3pt] \hline
$\delta\theta^\odot_{\tt gr}~[15^\circ]$, ~mas &
10.180&
3.366 & 
2.778 &
2.341 &
1.746 &
1.372 &
1.122 &
0.948 \\[3pt] \hline
$\delta\theta^\odot_{\tt gr}~[1^\circ]$, ~mas &
1.123 &
0.297 & 
0.238 &
0.195 &
0.140 &
0.107 &
0.085 &
0.071 \\[3pt] \hline
\end{tabular} 
\end{center}
\end{table}
\begin{figure}
\noindent
\vskip -50pt  
\begin{center} \hskip -45pt 
\rotatebox{90}{\hskip 20pt Deflection angle    
~$\log_{10}[\theta^\odot_{\tt gr}], ~[\mu{\rm as}]$} \hskip -45pt
\begin{minipage}[b]{.46\linewidth}
\vskip -25pt
\centering\psfig{figure=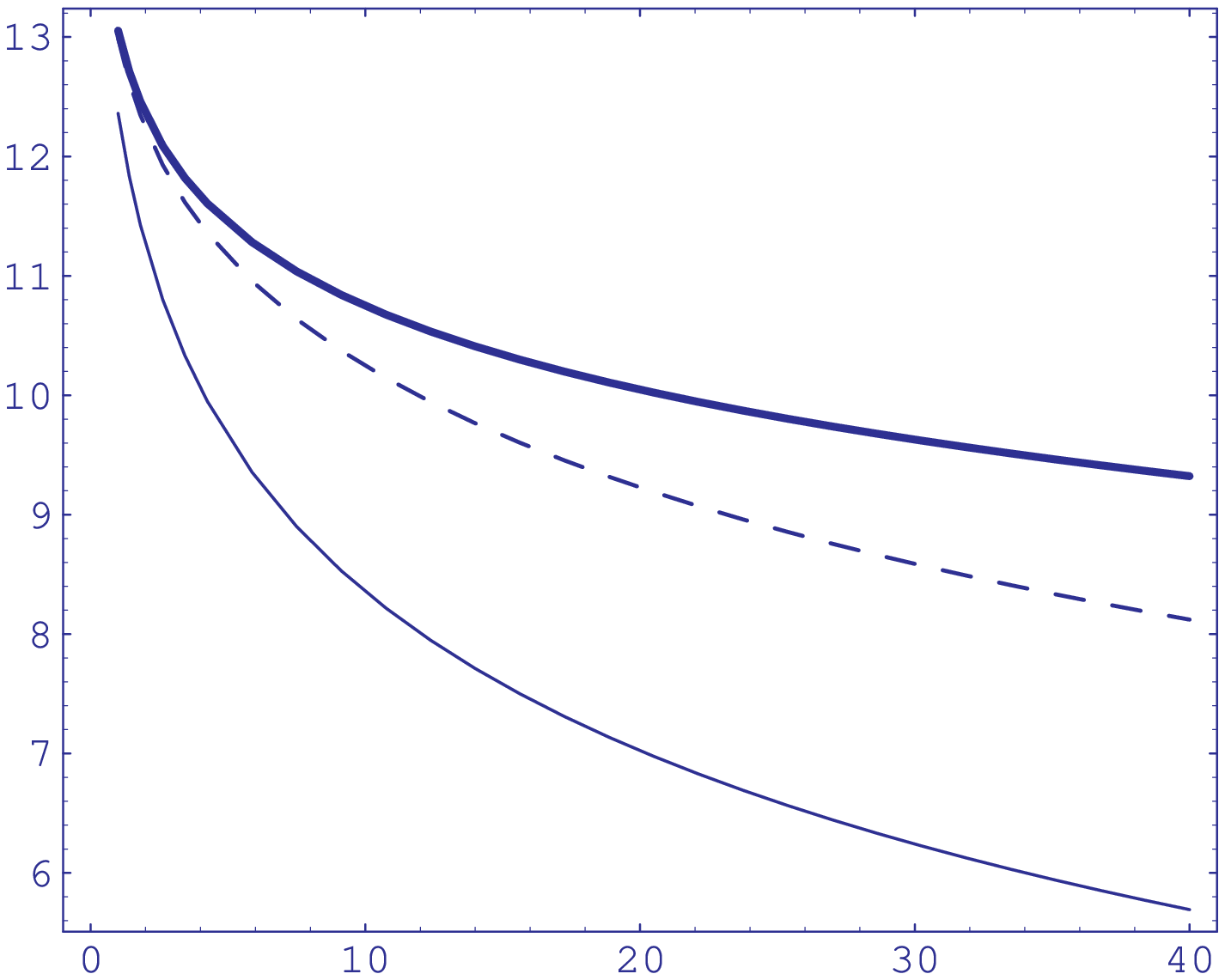,width=83mm,height=63mm}
\rotatebox{0}{\hskip 55pt  Sun-source separation angle,
~$\chi_{1\odot}$~[deg]}
\end{minipage}
\hskip 40pt
\rotatebox{90}{\hskip 20pt  
Deflection angle  ~$\log_{10}[\theta^\odot_{\tt gr}], ~[\mu{\rm as}]$}
\hskip -45pt
\begin{minipage}[b]{.46\linewidth}
\vskip -25pt
\centering \psfig{figure=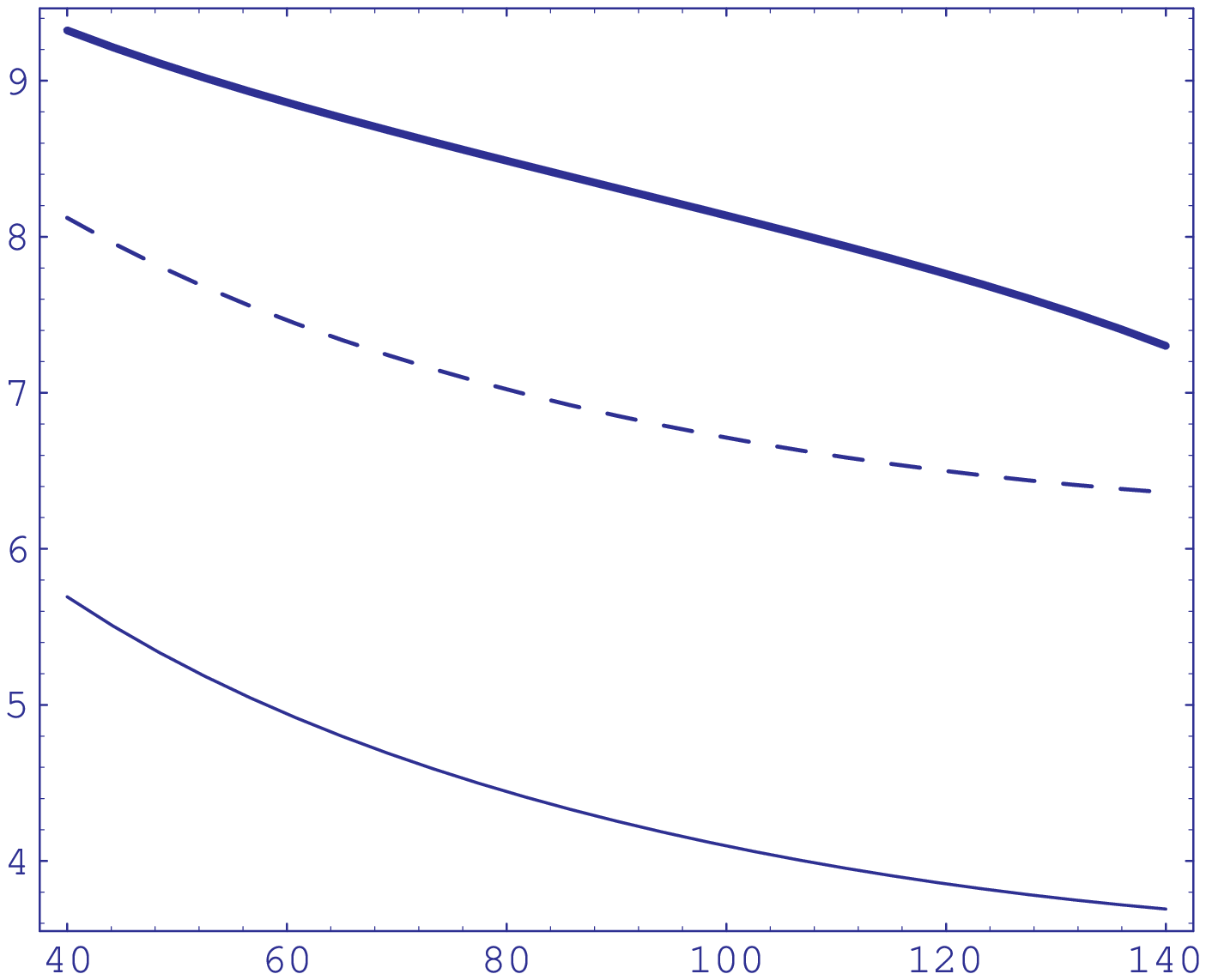,width=83mm,height=63mm}
\rotatebox{0}{\hskip 55pt  Sun-source separation angle,
~$\chi_{1\odot}$~[deg]}
\end{minipage}
\caption{Solar gravitational deflection of light. On all plots: the upper thick line is for the  absolute astrometric measurements, 
while the other two are for the differential astrometry. Thus, the  dashed line is for the observations over field of regard  of 
FoR$\,\,= 15^\circ$, the lower thick line is for FoR $ = 1^\circ$. 
 \label{fig:solar_defl}}
 \end{center}
\end{figure}


Figure \ref{fig:solar_defl} shows a qualitative presentation of the solar gravitational deflection. The upper thick line on both plots represents the absolute astrometric measurements, while the other two are for the differential astrometry. Thus, the middle dashed line is for the observations over the maximal field of regard  of the instrument FoR$\,\,= 15^\circ$, the lower thick line is for FoR $ = 1^\circ$. 


One can also account for the post-post-Newtonian (post-PN) terms (e.g.  $\propto G^2$) as well as the contributions due to  other PPN parameters \cite{Will_book93}.  Thus, in the weak gravity field approximation the total deflection  angle $\theta_{\tt gr}$ has an additional  contribution due the post-post-Newtonian terms in the metric tensor. 
For the crude estimation purposes this effect  could be given by the following expression \cite{post-PPN}:   
{}
\begin{equation}
\delta\theta_{\tt post-PN}=
\frac{1}{4}(\gamma+1)^2  
\left(\frac{2GM}{c^2d}\right)^2
\left(\frac{15\pi}{16}-1\right)
\left(\frac{1+\cos\chi}{2}\right)^2.
\label{eq:postPN}
\end{equation}

\noindent However, a quick look on the magnitudes of these terms  for the
solar system's bodies suggested that SIM astrometric data will be 
insensitive to the post-PN effects. The   post-PN effects   due to the  Sun are the largest among those in the solar system. However, even for the absolute astrometry with the Sun-grazing rays the post-PN terms were estimated to be of order $\delta\theta_{\tt post-PN}^\odot= 11~\mu$as. Note that the SIM solar avoidance angle is constraining  the Sun-source separation angle  as $\chi_{1\odot} \ge$ 45$^\circ$. The post-PN effect is inversely proportional to the square of the impact parameter, thus reducing the effect to $\delta\theta^\odot_{\tt post-PN}\le 4.9 $ nanoarcseconds at the edge of the solar avoidance angle. Therefore, the post-PN effects will not be accessible with SIM.
 
\vskip 20pt
\subsection{Gravitational Deflection of Light by Jupiter} 

Astrometric measurement with SIM  would have to account for the light bending by Jupiter \cite{Crosta-Mignard-2006,Fomalont-Kopeikin07}. One may obtain the expression, similar to Eq.~(\ref{eq:sun_defl}) for
the relativistic deflection of light by the Jovian gravitational
monopole in the following form:    {} 
\begin{equation}
\theta^J_{\tt gr}= (\gamma+1)~\frac{G}{c^2}
\frac{M_J}{r_J}~\frac{1+\cos\chi_{1J}}{\sin\chi_{1J}}=
0.924944\cdot\frac{1+\cos\chi_{1J}}{\sin\chi_{1J}}
~~~~\mu{\sf as},
\label{eq:deflec_jup}
\end{equation}
\noindent with $\chi_{1J}$ being Jupiter-source separation angle as seen by the interferometer at the distance $r_J$ from Jupiter. For the differential  observations one will have expression, similar to that Eq.~(\ref{eq:sun_def_defl}) for the Sun:   
{} 
\begin{equation}
\delta\theta^J_{\tt gr}=(\gamma+1)\frac{G}{c^2} ~
\frac{M_J}{r_J}\frac{\sin\frac{1}{2}(\chi_{2J}-\chi_{1J})}
{\sin\frac{1}{2}\chi_{1J}\cdot\sin\frac{1}{2}\chi_{2J}}=0.924944
\cdot\frac{\sin\frac{1}{2}(\chi_{2J}-\chi_{1J})}
{\sin\frac{1}{2}\chi_{1J}\cdot\sin\frac{1}{2}\chi_{2J}}~~~~\mu{\sf as}, 
\label{eq:deflec_diff_jup}
\end{equation}
  
\noindent where again $\chi_{1J}, \chi_{2J}$ are Jupiter-source
separation angles for the primary and secondary stars
correspondingly, $\chi_{2J}=\chi_{1J}+\frac{\pi}{12}$
(and $\chi_{2J}=\chi_{1J}+\frac{\pi}{180}$ for the narrow angle
astrometry).   The largest effect will come when SIM and Jupiter
are at the closest distance from each other   $\sim 4.2 ~$AU.
Jupiter's angular dimensions from the Earth' orbit for this
situation are  calculated to be ${\cal R}_J=23.24$ ~arcsec,  which correspond to a deflection angle of 16.419 mas.  Results for some $\chi_{1J}$ are given in the  Table \ref{tjes}. Note that for the light rays coming perpendicular to the ecliptic plane the Jovian deflection will be in the range: $\delta\alpha_{1J}\sim(0.7\--1.0)~\mu$as!

\begin{figure}[h]
\noindent  
\begin{center} \hskip -45pt 
\rotatebox{90}{\hskip 20pt Deflection angle    
~$\log_{10}[\theta^J_{\tt gr}], ~[\mu{\rm as}]$}
\hskip -45pt
\begin{minipage}[b]{.46\linewidth}
\vskip -25pt
\centering\psfig{figure=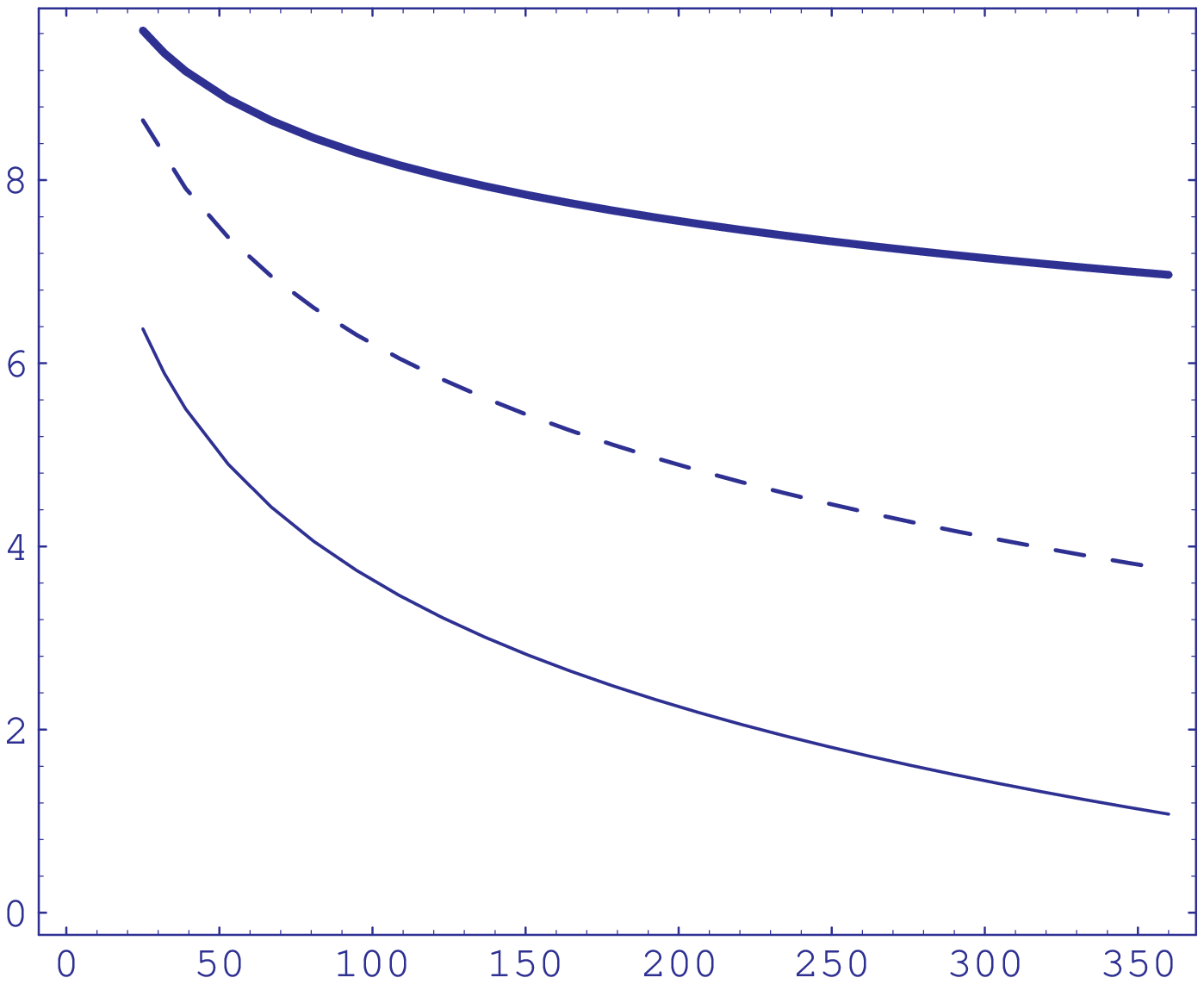,width=85mm,height=70mm}
\rotatebox{0}{\hskip 43pt  Jupiter-source separation angle,
~$\chi_{1J}$~[arcsec]}
\end{minipage}
\hskip 40pt
\rotatebox{90}{\hskip 20pt  
Deflection angle  ~$\log_{10}[\theta^J_{\tt gr}], ~[\mu{\rm as}]$}
\hskip -45pt
\begin{minipage}[b]{.46\linewidth}
\vskip -25pt
\centering \psfig{figure=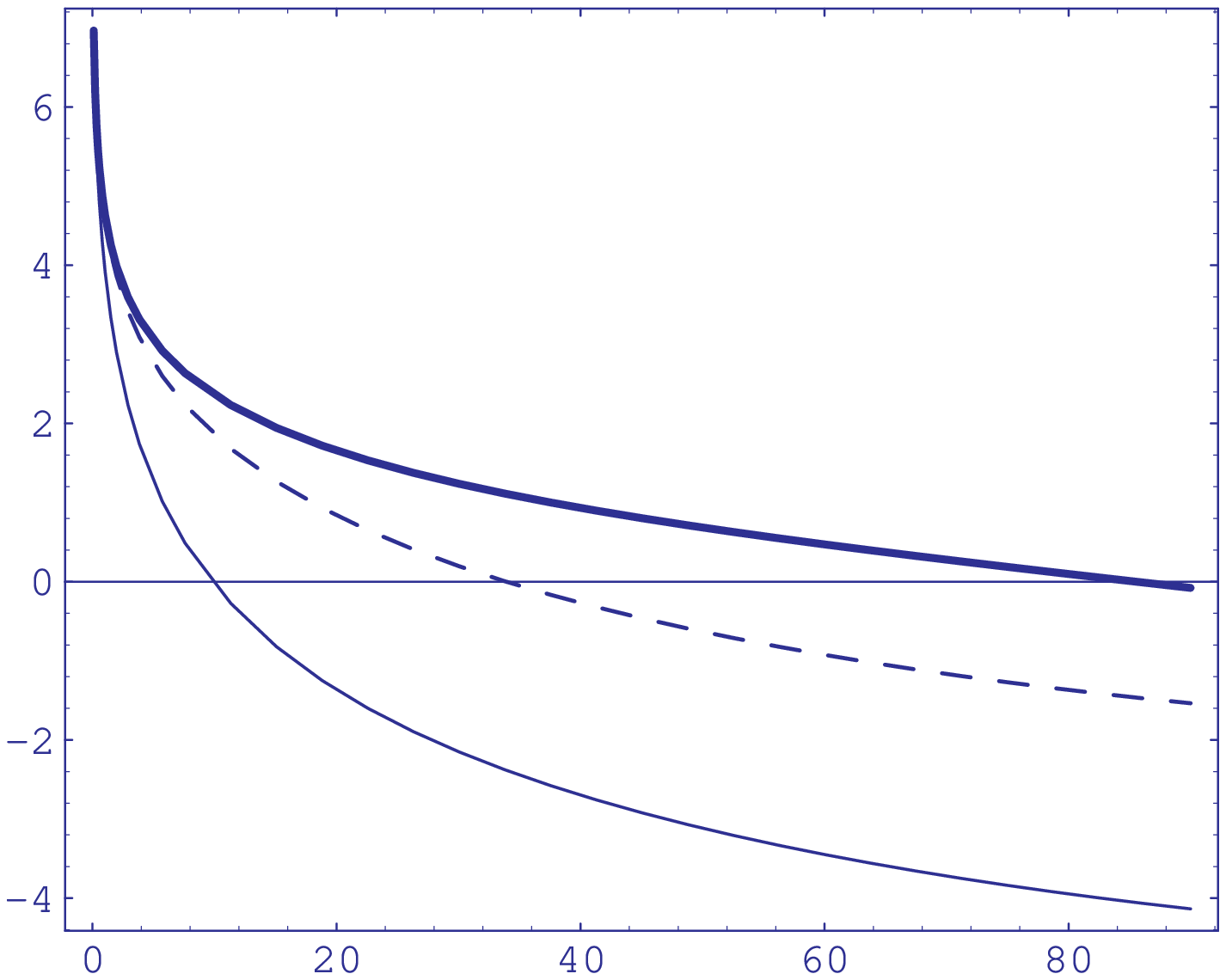,width=85mm,height=70mm}
\rotatebox{0}{\hskip 50pt  Jupiter-source separation angle,
~$\chi_{1J}$~[deg]}
\end{minipage}
   \caption{Jovian gravitational deflection of light. 
   \label{fig:jupiter_defl}}
 \end{center}
\end{figure}

A qualitative behavior of the effect of the gravitational
deflection of light by the Jovian gravity field is plotted in the 
Figure \ref{fig:jupiter_defl}. As in the case of the solar deflection, the upper thick line on both plots represents the absolute astrometric measurements, while the other two are for the differential astrometry (the dashed  line is for the observations over FoR$\,\,= 15^\circ$ and 
the lower thick line is for FoR $ = 1^\circ$). 


\vskip 10pt
\begin{table}[h]
\caption{Jovian gravitational monopole deflection  vs. the 
Jupiter-source sky separation angle  $\chi_{1J}$.}
\label{tjes}
\begin{center}
\begin{tabular}{|r|c|c|c|c|c|c|c|c|}
\hline
\rm Jovian   & \multicolumn{8}{c|}{ Jupiter-source
separation angles $\chi_{1J}, ~~$arcsec}  \\
\cline{2-9} 
\rm deflection 
&$23.24''$ &$26''$&$30''$&$60''$&$120''$&$180''$&$360''$
&$90^\circ$\\\hline
$~\theta^J_{\tt gr},~$mas &
16.419 &
14.676 & 
12.719 &
6.360 &
3.180 & 
2.120 &
1.060 &
$0.9~\mu$as\\ \hline
$\delta\theta^J_{\tt gr}[15^\circ],~$mas &
16.412 &
14.669 & 
12.712 &
6.352 &
3.173 & 
2.113 &
1.053 &
$0.2~\mu$as\\ \hline
$\delta\theta^J_{\tt gr}[1^\circ],~$mas &
16.313 &
14.570 & 
12.614 &
6.255 &
3.077 & 
2.019 &
0.964 &
$0.0~\mu$as\\ \hline
\end{tabular} 
\end{center}
\end{table}

\subsection{Gravitational Deflection of Light by the Earth} 

The deflection of light rays by the Earth's gravity field may also be of
interest. The expressions, describing  the relativistic deflection
of light by the Earth' gravitational monopole are given below:   
{} 
\begin{equation}
\theta^\oplus_{\tt gr}= (\gamma+1)~\frac{G}{c^2}
\frac{M_\oplus}{r_\oplus}~\frac{1+\cos\chi_{1\oplus}}{\sin\chi_{1\oplus}}=
0.2446\cdot\frac{1+\cos\chi_{1\oplus}}{\sin\chi_{1\oplus}}
~~~~\mu{\sf as},
\label{eq:deflec_earth}
\end{equation}
\noindent with $\chi_{1\oplus}$ being the Earth-source separation angle as seen by the interferometer at the distance $r_\odot$ from the Earth. Relation for the differential astrometric measurements was
obtained in the form:  
 
\begin{equation}
\delta\theta^\oplus_{\tt gr}=(\gamma+1)\frac{G}{c^2} ~
\frac{M_\oplus}{r_\oplus}
\frac{\sin\frac{1}{2}(\chi_{2\oplus}-\chi_{1\oplus})}
{\sin\frac{1}{2}\chi_{1\oplus}\cdot\sin\frac{1}{2}\chi_{2\oplus}}
=0.2446\cdot\frac{\sin\frac{1}{2}(\chi_{2\oplus}-\chi_{1\oplus})}
{\sin\frac{1}{2}\chi_{1\oplus}\cdot\sin\frac{1}{2}\chi_{2\oplus}}
~~~~\mu{\sf as}, 
\label{eq:deflec_diff_earth}
\end{equation}
  
\noindent where, as before, $\chi_{1\oplus}, \chi_{2\oplus}$ are the
Earth-source separation angles for the primary and secondary stars
correspondingly, $\chi_{2\oplus}=\chi_{1\oplus}+\frac{\pi}{12}$ (and $\chi_{2\oplus}=\chi_{1\oplus}+\frac{\pi}{180}$ for the narrow angle astrometry).   The largest effect will come when SIM and the Earth are at the closest distance, say at the end of the first half of the  first year of the mission, $r_\oplus=0.05$ AU. The Earth's angular dimensions being measured from the spacecraft from that distance are  calculated to be ${\cal R}^{\tt SIM}_\oplus=175.88401$ arcsec, which correspond to a deflection angle of ~ 573.75 $\mu$as. The summary of the deflection angles for sevral $\chi_{1\oplus}$ are given in the Table \ref{tab:sim_hip}. 
 

\begin{table}[h]
\begin{center}
\caption{Solar relativistic deflection angle as a function of the
Earth-source  separation angle. 
\label{tab:sim_hip}}\vskip 10pt
\begin{tabular}{|r|c|c|c|c|c|c|c|}
\hline

\tt SIM    & \multicolumn{7}{c|}{$\chi^{\tt SIM}_{1\oplus},$ ~arcsec}   \\
\cline{2-8}
\tt mission & 175.88 & 200 & 360 &
$1^\circ$ & $5^\circ$ & $10^\circ$ & $15^\circ$\\\hline 
$~~\theta_{1\oplus},~\mu$as &
573.8 &
504.7 & 
280.3 &
28.0 &
5.6 & 
2.8 &
1.9 \\ \hline
$\delta\theta_{1\oplus}[15^\circ],~\mu$as &
571.9 &
502.7 & 
278.5 &
26.3 &
4.2 & 
1.7 &
1.0 \\ \hline
$\delta\theta_{1\oplus}[1^\circ],~\mu$as &
547.0 &
478.0 & 
254.8 &
14.0 &
0.9 & 
0.3 &
0.1 \\ \hline
\end{tabular} 

\end{center}
\end{table}

\begin{figure}[h]
\noindent  
\begin{center} \hskip -45pt 
\rotatebox{90}{\hskip 20pt Deflection angle    
~$\log_{10}[\theta^\oplus_{\tt gr}],~[\mu{\rm as}]$}
\hskip -45pt
\begin{minipage}[b]{.46\linewidth}
\rotatebox{0}{\hskip 52pt At the distance of 0.05 AU from the Earth}
\vskip -30pt
\centering\psfig{figure=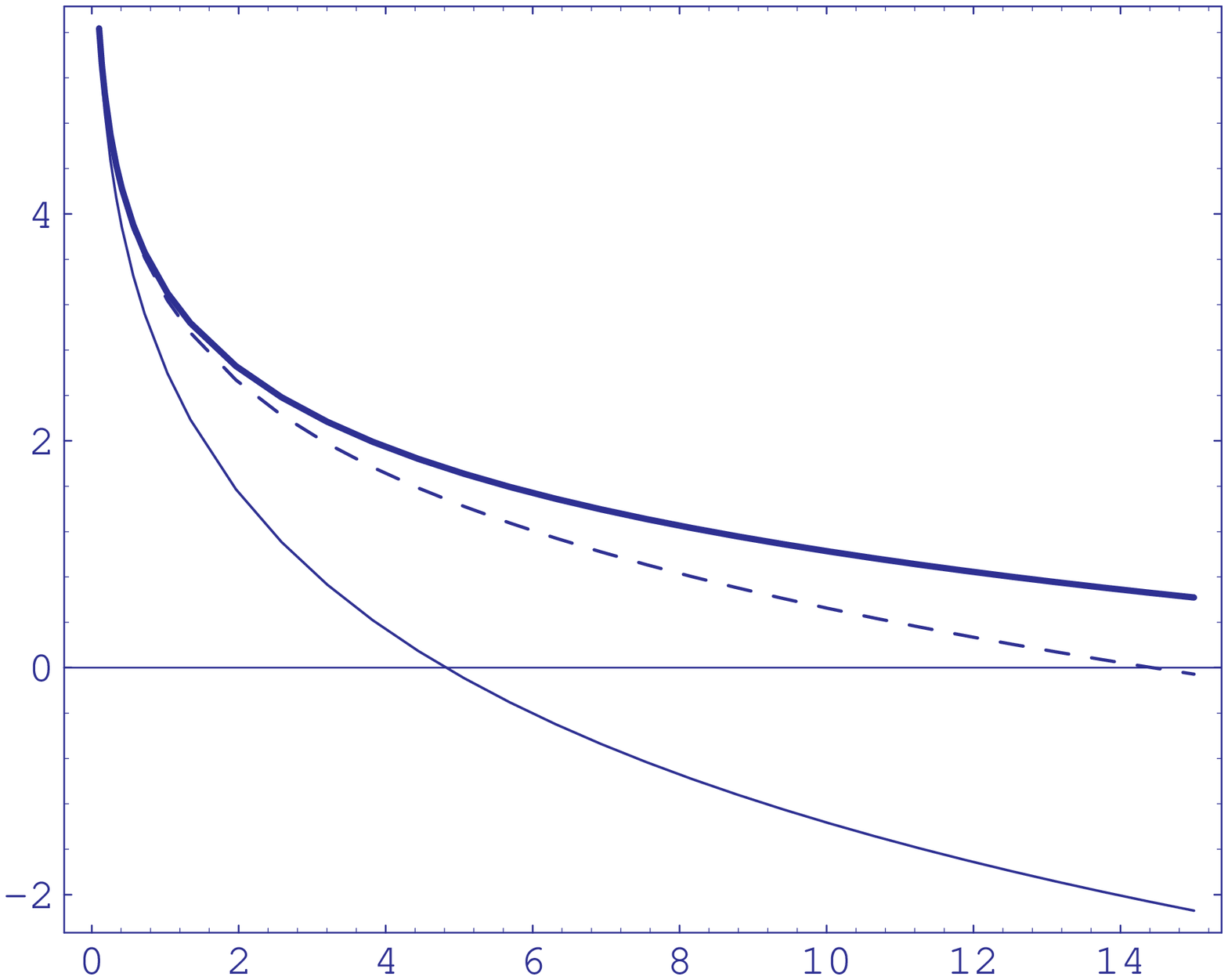,width=85mm,height=70mm}
\rotatebox{0}{\hskip 57pt  Earth-source separation angle,~$\chi_{1\oplus}$~[deg]}
\end{minipage}
\hskip 40pt
\rotatebox{90}{\hskip 20pt  
Deflection angle  ~$\log_{10}[\theta^\oplus_{\tt gr}], ~[\mu{\rm as}]$}
\hskip -45pt
\begin{minipage}[b]{.46\linewidth}
\rotatebox{0}{\hskip 55pt   At the distance of 0.5 AU from the Earth}
\vskip -30pt
\centering \psfig{figure=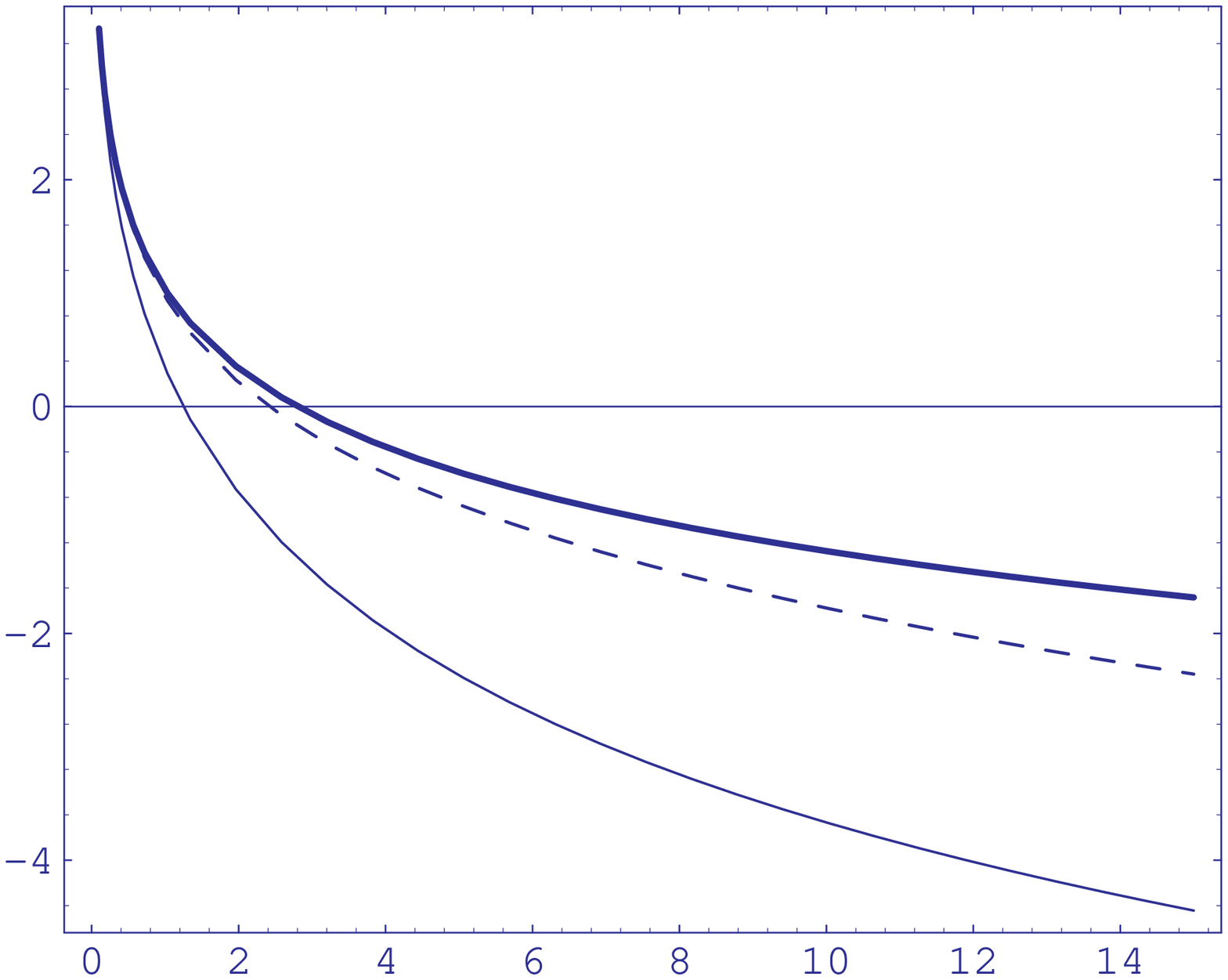,width=85mm,height=70mm}
\rotatebox{0}{\hskip 57pt  Earth-source separation angle,~$\chi_{1\oplus}$~[deg]}
\end{minipage}
     \caption{Gravitational deflection of light in the proximity of the Earth. 
      \label{fig:earth_delf}}
 \end{center}
\end{figure}
 
Figure \ref{fig:earth_delf} shows the expected variation in the magnitude of the Earth' gravity influence as mission progresses. The left plot presents results for the end of the first half of the year of the mission, when the spacecraft is at the distance of 0.05 AU from the Earth (the drift rate is 0.1 AU per year). The plot on the right side is for the end of the 5-th year of the mission, when SIM is at 0.5 AU from Earth.

\subsection{Constraints Derived From the Monopole Deflection of Light}
\label{sec:constr}

While analyzing the solar gravity field's influence on the future
astrometric observations with SIM, we found several interesting
situations, that may potentially put an additional navigational
requirements. In this section we will consider these situations in a
more detailed way.

To measure gravitational deflection of light with an accuracy of $\Delta \theta_0=\Delta k ~\mu{\rm as}$, one needs to precisely determine the value of impact parameter of photon's trajectory with respect to the deflector. 
As before, we will present two types of necessary expressions, 
namely for absolute and  differential observations.

In the case of absolute astrometry we use expression for the deflection angle $\theta^B_{\tt gr}$ from Eq.~(\ref{eq:defl_chi-ind}) and present it as $\theta^B_{\tt gr}= ({\mu_B}/{d})(1+\cos\chi_B)$, where again $d=r_B\sin\chi_B$ and quantity $\mu_B=2GM_B/c^{2}$ being the relativistic gravitational radius of the body at question. One may ask a question -- what uncertainty in the knowledge of the impact parameter, $\Delta d$, will result in the astrometric error of $\Delta \theta_0$? The answer is given the following expression   
{}
\begin{equation}
\Delta d_B= \Delta \theta_0\,
\,\frac{r^2_B}{\mu_B}\cdot
 \frac{\sin^2\chi_{1B}}{1+\cos\chi_{1B}}.
\label{eq:impact_parameter}
\end{equation}
\noindent The corresponding result for differential observations may be obtained with the help of Eq.~(\ref{eq:diff_opd}) as: 
{}
\begin{equation}
\Delta d^{\tt diff}_B= \Delta \theta_0\,
\,\frac{r^2_B\sin^2\chi_{1B}}{2\mu_B}\cdot
\Big[1+\tan\frac{\chi_{1B}}{2}\cdot
\cot\frac{1}{2}(\chi_{2B}-\chi_{1B})\Big].
\label{eq:impact_parameter_diff}
\end{equation}
\noindent Similarly, the uncertainty in determining the barycentric distance $r_B$ is obtained from Eq.~(\ref{eq:defl_chi-ind}) leading to expression: 
{}
\begin{equation}
\Delta r_B= \Delta \theta_0 \,\frac{r^2_B}{\mu_B}\cdot
 \frac{\sin\chi_{1B}}{1+\cos\chi_{1B}}.
\label{eq:bar_dist}
\end{equation}
Note that, when differential observations are concerned, uncertainty in barycentric position  $\Delta r^{\tt diff}_B$ does not produce new constraints significantly different from those derived from  Eq.~(\ref{eq:bar_dist}).  Looking at the results presented in the Table~\ref{tab6}, one may see that for an accuracy of $\Delta \theta_0=1~\mu$as our estimates require the knowledge of the solar impact parameter with the accuracy of $\sim 0.4$ ~km (grazing rays), that for Jupiter with the accuracy of $\sim 4 ~$km and other big planets with the accuracy of about  10 ~km.  Table~\ref{tab_best} shows a comparison of these derived requirements on the barycentric positions of the solar system's bodies  with the accuracy of their current determination. 
 
\begin{table}
\begin{center}
\caption{Required accuracy of   
barycentric positions and impact parameters for astrometric
observations with accuracy of 1 $\mu$as. The Earth is taken at the distance of 0.05 AU from the spacecraft; moon's position accuracy is for the geocentric frame. } 
\label{tab6}\vskip 10pt

\begin{tabular}{|r|c|c|c||c|c|c|c|} \hline

Solar & \multicolumn{3}{c||}{Required knowledge: grazing rays}  
& \multicolumn{4}{c|}{Required knowledge: differential astrometry}
\\\cline{2-4}\cline{5-8}
system's & Distance, & \multicolumn{2}{c||}{Impact parameter} & 
\multicolumn{2}{c|}{Impact param. [15$^\circ$]}  &  
\multicolumn{2}{c|}{Impact param. [1$^\circ$]} \\\cline{3-4}\cline{5-8} 
object & $\sigma_{r_B}, $~km & $\sigma_{d_B}, $~km &
$\sigma_{d_B}, $~mas & $\sigma_{r_B}$,~km & $\sigma_{r_B}$,~mas&
$\sigma_{r_B}$,~km & $\sigma_{r_B}$,~mas \\[2pt]\hline\hline

Sun     & 85.45 & 0.39& 0.55 &  0.40 & 0.55  & 0.50 & 0.69 \\

Sun   at 45$^\circ$  & 1.5$\times10^4$ & 7.6$\times10^3$
        & 10$''$.49 & 3.81$\times10^4$  
        & 52$''$.53 & 4.45$\times10^5$ & 613$''$.66\\ 

Moon    & 2.8$\times10^5$ & 67.14 &  1$''$.85 
        & 67.17  & 1$''$.85  &68.00 & 1$''$.88 \\

Mercury &1.1$\times10^6$& 29.39 & 66.16  & 29.41 
        & 66.16 & 29.45 & 66.25\\

Venus   & 8.4$\times10^4$ & 12.18 & 61.00  & 12.28 & 61.20  
        & 12.38 & 62.70 \\

Earth-Moon  & 1.3$\times10^4$ & 11.14 & 306.55  
            & 11.15 & 307.47 & 11.66 & 321.54\\

Mars    & 6.8$\times10^5$ & 29.29 & 77.11 & 29.30  & 77.14  
        & 29.37 & 77.33\\ 

Jupiter & 3.8$\times10^4$ & 4.31 & 1.42 & 4.32 & 1.42& 4.34 & 1.42\\

Jupiter at 30$''$ & 4.9$\times10^4$ & 7.14 
        & 2.34 & 7.20 & 2.36 & 7.25 & 2.38\\ 

Saturn  & 2.2$\times10^5$ & 10.32 & 1.66 & 10.34   
        & 1.66 & 10.36 & 1.66\\

Uranus  & 1.2$\times10^6$ & 11.27 & 0.86 & 11.28   
        & 0.86 & 11.29 & 0.86\\

Neptune & 1.7$\times10^6 $& 10.04 & 0.47 & 10.04  
        & 0.47 & 10.04 & 0.47 \\

Pluto    & 2.0$\times10^9$ & 1\,133.92 & 40.7
         & 1\,133.93   & 40.7 & 1133.96& 40.7\\\hline

\end{tabular} 
\end{center} 
\end{table}
 
One may see that the present accuracy of knowledge of the inner planets' positions  from the Table \ref{tab_best} is given by the radio observations and it is  even better than the level of relativity requirements given in  the Table \ref{tab6}. However, the positional accuracy for the outer planets is below the required level. The SIM observation program should include the astrometric studies  of the outer planets in order to minimize the errors in their positional accuracy determination. Thus, in order to get the radial uncertainty in Pluto's ephemeris with accuracy below 1000 ~km, it is necessary only 4 measurements of Pluto's position, taken sometime within a week of the stationary points,  spread over 3 years. Each measurement could be taken with an accuracy of about $200 ~\mu$as, as suggested by \cite{Standish95}. Additionally, one will have to  significantly lean on the radio observations in order to conduct the reduction of the optical data with an accuracy of a few $\mu$as.  For this reason one will have to use  the precise catalog of the radio-sources and to study the problem of the  radio and  optical reference frame ties \cite{Standish95,Standish_etal95,Folkner94,Ma-1998}.

The estimates, presented here were given for static gravitational field. Analysis of a real experimental situation should  consider  a non-static gravitational environment of the solar system  and should include the description of light propagation in a different reference frames involved in the experiment \cite{Kopeikin-Schafer-1999}. Additionally, the  observations will be affected by the relativistic orbital dynamics of the spacecraft \cite{Turyshev98}. 

\begin{table}
\begin{center}
\caption{The best known accuracies of barycentric positions and masses for the solar system's objects \cite{Yoder95,Pitjeva-2005,Folkner-Williams-Boggs-2008}.
} 
\label{tab_best} 
\vskip 10pt 
\begin{tabular}{|r|c|c|c||c|} \hline

Solar   
& \multicolumn{3}{c||}{Knowledge of barycentric position}
& Knowledge of  \\\cline{2-4}
system's &  \multicolumn{2}{c|}{Best known,} & Method used 
& planetary masses,   \\\cline{2-3}  
object & $\sigma_{r_B}, $~km & $\sigma_{r_B}, $~mas & 
for determination & 
$\Delta M_B/M_B$   \\[2pt]\hline\hline

Sun    &  362/725 & $0''.5/1''.0$&  
        Optical meridian transits  &  3.77$\times10^{-10}$   \\


Moon    & 27 ~cm & 7.4 $\mu$as & LLR, 1995 
& 1.02$\times10^{-6}$   \\

Mercury & 1 & 2.25& Radar ranging & 4.13$\times10^{-5}$   \\

Venus   & 1  & 4.98& Radar ranging & 1.23$\times10^{-7}$   \\

Earth   & 1 & 27.58 & Radar ranging  & TBD$\times10^{-6}$   \\

Mars    & 1 & 2.63& Radar ranging & 2.33$\times10^{-6}$   \\ 

Jupiter & 30 & 9.84& Radar ranging & 7.89$\times10^{-7}$  \\

Saturn  & 350 & 56.24& Optical astrometry & 2.64$\times10^{-6}$   \\

Uranus  & 750 & 57.00 & Optical astrometry & 3.97$\times10^{-6}$   \\

Neptune & 3\,000 & 141.67& Optical astrometry & 2.19$\times10^{-6}$   \\

Pluto/Charon    & 20\,000   & 717.40 & Photographic astrometry 
        & 0.014  \\\hline

\end{tabular} 
\end{center} 
\end{table} 

\section{Deflection of light by higher gravity multipoles}
\label{sec:high_multipoles}

In order to carry out a complete analysis of the relativistic light deflection one should account for other possible terms in the expansion (\ref{eq:deflec0}) that may potentially contribute to this effect. These terms are due to non-sphericity and non-staticity of the body's gravity field \cite{Kopeikin-1997}.  Here we will consider several of them, namely those due to higher gravitational multipoles of the celestial bodies.  

\subsection{Gravitational quadrupole deflection of light}

Although a complete three-dimensional deflection of light must be considered for a real experiment, for the purposes of this paper, we consider only two dimensional case. In this case, quadrupole term may be given  as \cite{Turyshev98,Klioner-1991,Kopeikin-Makarov-2007}:
{}
\begin{equation}
\theta_{J_2}= \frac{1}{2}(\gamma+1) \frac{4G M_B}{c^2{\cal R}_B}J_{2B}\Big(1-s_z^2-2u_z^{\hskip 2pt 2} \Big)
\left({{\cal R}_B\over d}\right)^3,
\label{eq:quad}
\end{equation}
\noindent where $J_2$ is the second zonal harmonic of the body under question, $\vec{s}=(s_x,s_y,s_z)$ is the unit vector in the direction of the light ray propagation and vector $\vec{d}=d(u_x,u_y,u_z)$ is the impact parameter. A similar expression may also be obtained for differential observations. For estimation purposes, this formula may be given as follows ($d=r_B\sin\chi_B$):
{}
\begin{equation}
\delta\theta_{J_2}\approx \frac{4\mu_B  J_{2B}{\cal R}^2_B}
{r^3_B} ~\Big[\,\frac{1}{\sin^3\chi_{1B}}-\,\frac{1}{\sin^3\chi_{2B}} \Big].
\label{eq:quad_diff}
\end{equation}
\noindent The corresponding effects for the deflection of light by the quadrupole mass moments within the planets of the solar system are given in the Table \ref{tab11}. The effect depends on a number of different instantaneous geometric parameters defining the mutual orientation of the vector of the light propagation, position of the planet in orbit, the orientation of the axes defining it's figure, etc. A mission-independent modeling in the static gravitational regime has been done \cite{Crosta-Mignard-2006}; the modeling of the dynamic regime was presented in \cite{Kopeikin-Makarov-2007}. An effort to develop a SIM-specfic model is well justified.   

The quadruple deflection of light depends on the third power of the inverse impact parameter with respect to the deflecting body, Eq.~(\ref{eq:quad}). SIM will measure this effect directly for many celestial bodies. At the expected level of accuracy the  knowledge of jovian atmosphere, the magnetic field  fluctuations, etc.,  may contribute to the errors in the experiment  \cite{TreuhaftLowe91}. A detailed study of these effects is given in \cite{Kopeikin-Fomalont-02}.


\begin{table}[t]
\begin{center}
\caption{Higher gravitational coefficients for solar system bodies ({\tt http://nssdc.gsfc.nasa.gov/planetary/factsheet/}).}  
\label{tab14}\vskip 10pt
\begin{tabular}{|r|c|c|c|} \hline
Solar system's & $J^B_2,$ & $J^B_4,$ & 
$J^B_6,$\\
object & $ \times ~10^{-6}$ & $\times ~10^{-6}$ & 
$\times ~10^{-6}$\\\hline\hline 

Sun & $0.17\pm0.017$& | & |  \\

Sun at 45$^\circ$ & | & |  & |  \\ 

Moon & 202.2  & $-0.1$   & |   \\

Mercury & 60.  &  | &  |  \\

Venus & 4.5  & $-2.1$ &  |  \\

Earth &  1\,082.6  & $-1.6$ &  0.5 \\
 
Mars & 1\,960.45 & | & | \\

Jupiter & 14\,738$\pm$1 & $-$587$\pm$5 & 34$\pm$50\\ 

Saturn & 16\,298$\pm 50$ & $-$915$\pm$80 & $103.0$\\ 

Uranus  & 3\,343.43  & |   & |   \\

Neptune  & 3\,411.  & | & |  \\
 
Pluto    &  |  &  | & |  \\\hline

\end{tabular}
\end{center}
\end{table} 
\begin{table}[h]
\begin{center}
\caption{Relativistic deflection of light by the planetary quadrupole mass moments for solar system bodies.} 
\label{tab11}  \vskip 10pt 
\begin{tabular}{|r||c|c|c||c|c|} \hline
Solar system' & $J^B_2,$ & $\theta^B_{J_2}, $ &
$d^{\tt crit}_{{J_2}}$   & $\delta\theta^B_{J_2}[15^\circ],$ & 
$\delta\theta^B_{J_2}[1^\circ],$\\
object & $ \times ~10^{-6}$ & $\mu$as &
 & $\mu$as & 
$\mu$as\\\hline\hline 

Sun & $0.17\pm0.017$& 0.3 & | & 0.3  & 0.3  \\


Moon & 202.2  & 2$\times 10^{-2}$ &   | 
              & 2$\times 10^{-2}$ & $2\times 10^{-2}$    \\

Mercury & 60.  & 5$\times 10^{-3}$  &  |&  | &  |   \\

Venus & 4.5  & 2$\times10^{-3}$ &  | & | &  |  \\

Earth &  1\,082.6  & 0.6  & | & 0.6  &  0.6 \\
 
Mars & 1\,960.45 & 0.2 & | & 0.2 & 0.2 \\\hline

Jupiter &  14\,738$\pm$1  & 242.0 &  98$''$.12 \--- 144$''$.81  & 242.0  
& 242.0  \\
  &     &  &  6.23 ${\cal R}_J$  &    &  \\\hline


Saturn &  16\,298$\pm$ 50  & 94.6 &  35$''$.62 \--- 43$''$.93 & 94.6 
& 94.6 \\ 
  &      & & 4.56 ${\cal R}_S$   &  &   \\\hline 

Uranus  & 3\,343.43  & 7.3 &  3$''$.25 \--- 3$''$.61 & 7.3   & 7.3   \\
        &    &  & 1.94 ${\cal R}_U$  &     &    \\\hline

Neptune  & 3\,411.   & 8.5 & 2$''$.23 \--- 2$''$.42 & 8.5 & 8.5  \\
  &    &  &  2.04~${\cal R}_N$ &  &    \\\hline
 
Pluto    &  |  &  |  & | &  | & |   \\\hline

\end{tabular}
\end{center}
\end{table}

Table~\ref{tab11} presents estimates of the magnitudes of the relativistic deflection of light by the planetary quadrupole mass moments for solar system bodies. Based on the sizes of these effects, one would have to account for the quadrupole component of the gravity fields when observations will be conducted in the vicinity of the outer planets. In addition, the influence of the  higher harmonic may be also of interest. Thus, Table \ref{tab11} shows the estimates of some higher gravitational multipole moments of Jupiter and Saturn.  We will discuss the deflection by the $J_2$ and $J_4$ coefficients of the jovian gravity field in terms of Jupiter-source separation angle $\chi_{1J}$. An  expression, similar to that of Eq.~(\ref{eq:deflec_jup}) for the monopole deflection, may be given as:
{} 
\begin{equation}
\theta^{\tt max}_{J_2}={3.46058\times 10^{-10}}~
\frac{1}{\sin^3\chi_{1J}} ~~\mu{\sf as}.
\label{(2.16)}
\end{equation}

\noindent 
Jupiter's angular dimensions from the Earth are  calculated to be ${\cal R}_J=23.24 ~$arcsec, which correspond to a deflection  angle of $242 ~\mu$as. The deflection on the multipoles  for some   $\chi_{1J}$ is given  in the Table \ref{tab12}. 


\begin{table}[h]
\begin{center}
\caption{Deflection of light by the Jovian higher gravitational
coefficients.} 
\label{tab12}\vskip 10pt
\begin{tabular}{|r|c|c|c|c|c|c|c|}
\hline

\rm Jovian   & \multicolumn{7}{c|}{$\chi_{1J}$, arcsec}   
\\ \cline{2-8} 
\rm deflection &$23''.24$ &26$''$&30$''$&35$''$&40$''$&50$''$
&120$''$\\\hline\hline
$\theta^J_{J_2}, ~\mu$as &
242 &
173 & 
112 &
71 &
47 & 
24 &
1.8 \\ \hline

$\delta\theta^J_{J_2}[15^\circ], ~\mu$as &
242 &
173 & 
112 &
71 &
47 & 
24 &
1.8  \\ \hline\hline

$\theta^J_{J_4}, ~\mu$as &
9.6 &
5.5 & 
2.7 &
1.3 &
0.6 & 
0.2 &
0.0  \\ \hline
\end{tabular} 
\end{center}
\vskip 10pt
\begin{center}
\caption{Deflection of light by the Saturnian higher gravitational
coefficients.} 
\label{tab12sat}\vskip 10pt
\begin{tabular}{|r|c|c|c|c|c|c|c|}
\hline

\rm Saturnian   & \multicolumn{7}{c|}{$\chi_{1S}$, arcsec} \\
\cline{2-8} 
\rm deflection &$9''.64$ &12$''$&15$''$&20$''$&25$''$&30$''$&35$''$
\\\hline\hline
$\theta^S_{J_2}, ~\mu$as &
94.7 &
49.1 & 
25.1 &
10.6 &
5.4 & 
3.1 &
2 \\ \hline
$\delta \theta^S_{J_2}[15^\circ], ~\mu$as &
94.7 &
49.1 & 
25.1 &
10.6 &
5.4 & 
3.1 &
2 \\ \hline\hline

$\theta^S_{J_4}, ~\mu$as &
5.3 &
1.8 & 
0.6 &
0.1 &
| & 
| &
| \\ \hline

\end{tabular} 
\end{center}
\end{table}

Similar studies are important for Saturn (see \cite{Kopeikin-Makarov-2007} for details). In terms of the Saturn-source separation angle $\chi_{1S}$ the saturnian quadrupole  deflection mat be estimated with the help of the following expression: 
{}
\begin{equation}
 \theta^{\tt max}_{J_2}=9.66338\times 10^{-12}~
\frac{1}{\sin^3\chi_{1S}} ~~\mu{\sf as}.
\label{eq:guad_sat}
\end{equation}
The Saturn's angular dimensions  from the Earth' orbit  are  calculated
to be ${\cal R}_S=9.64 ~$arcsec, which correspond to a deflection  angle of $94.7~\mu$as.  The corresponding estimates for the deflection angles are presented in the Table \ref{tab12sat}.

As a result, for  astronomical observations with  accuracy of about   $1 ~\mu{\rm as}$, one will have to account for the quadrupole gravitational fields of the Sun, Jupiter, Saturn, Neptune, and Uranus. In addition, the influence of the higher harmonics may be of interest. For example some of the moments for Jupiter and Saturn are given in the Table~\ref{tab14}. 

Higher multipoles   may also influence the astrometric observations taken close to these planets. Thus, for both Jupiter and Saturn the  rays, grazing their surface, will  be deflected by the fourth zonal harmonic $J_4$ as follows: 
$\delta\theta^J_{J_4}\approx  9.6 ~\mu$as, $\delta\theta^S_{J_4}\approx  5.3 ~\mu$as. In addition, the contribution of the $J_6$ for Jupiter and Saturn will
deflect the grazing rays  on the angles $\delta\theta^J_{J_6}\approx 0.8 ~\mu$as, $\delta\theta^S_{J_6}\approx 0.6 ~\mu$as. The contribution of $J_4$ is decreasing with the distance from the body as $d^{-5}$ and contribution of $J_6$  as $d^{-7}$. As a result the deflection angle will be less then  $1 ~\mu$as when $\hskip 2pt d>1.6 ~{\cal R}$, where ${\cal R}$ is the radius of the planet.

Using Eq.~(\ref{eq:quad}), one can derive the expression for the critical distance $d^{\tt crit}_{J_2}$ for the  astrometric observations in the regime of quadrupole deflection of light with accuracy of $\Delta \theta_0= \Delta k ~\mu$as. Indeed, approximating this equation as $\theta_{J_2}\simeq ({4\mu_B}/{d^3})J_2{\cal R}^2$ and solving it for $d$ one obtaines the following result:  
{}
\begin{equation}
d^{\tt crit}_{J_2} = {\cal R}_B \Big[\frac{4\mu_B}{{\cal R}_B}  
\frac{J^B_2}{\Delta\theta_0}\Big]^\frac{1}{3}.
\label{eq:quad_crit_abs}
\end{equation}
\noindent 
The  critical distances for the relativistic quadrupole  deflection of
light by the solar system's bodies   for the case of $\Delta
k=1$ presented in the Table \ref{tab11}.

\subsection{Gravito-Magnetic Deflection of Light}

Besides the gravitational deflection of light by the monopole and the quadrupole components of the  static gravity field of the bodies, the light ray  trajectories will  also  be affected by the non-static contributions from this field.  It is easy to demonstrate  that a rotational motion of a  gravitating  body contributes to the total curvature of the space-time generated by this same body. This contribution produces an additional deflection of light rays \cite{Klioner-1991,Ciufolini-etal-2003,Kopeikin-Mashhoon-2003}  on the
angle 
\begin{equation}
\delta\theta_{\vec{\cal S}}=\frac{1}{2}(\gamma +1)
\frac{4G}{c^3d^3}\vec{\cal S}  (\vec{s}\cdot\vec{d}),
\label{eq:defl_spin}
\end{equation} 
\noindent where ${\vec{\cal S}}$  is the body's angular momentum. 

The most significant contributions of gravito-magnetic deflection of light by the bodies of the solar system are the following ones: i) the solar deflection amounts to $\delta\theta^\odot_{\vec{\cal S}}=\pm(0.7~-1.3)\mu$as (the first term listed is for a uniformly rotating Sun; the second is for the Dicke's model \cite{Dicke-1974}); ii) jovian rotation contributes $\delta\theta^J_{\vec{\cal S}}=\pm0.2~\mu$as;  and iii) Saturn's rotation $\delta\theta^{Sa}_{\vec{\cal S}}=\pm0.04~\mu$as.  Thus, depending on the model for the solar interior,  
solar rotation  may produce a noticeable contribution for the grazing rays.  The estimates of magnitude of deflection of light ray's trajectory, caused by the rotation of gravitating  bodies demonstrate that for precision of observations of $1 ~\mu$as it is sufficient to account for influence of the Sun and Jupiter only.

The relativistic gravito-magnetic deflection of light has never been directly tested before. However, because of the smallness of the magnitudes of corresponding  effects in the solar system and SIM's operational mode that limits the viewing angle for a sources as $\chi_{1\odot}\ge45^\circ$, SIM will not be sensitive to this effect. 

\section{Astrophysics Investigations with SIM} 
\label{sec:astrophys}

\subsection{Astrometric Test of General Relativity}
\label{sec:PPN-gamma} 

The Eddington parameter $\gamma$ in Eq.~(\ref{eq:metric}), whose value in general relativity is unity, is perhaps the most fundamental PPN parameter \cite{Will-lrr-2006-3}, in that $\frac{1}{2}(1-\gamma)$ is a measure, for example, of the fractional strength of the scalar gravity interaction in scalar-tensor theories of gravity \cite{Damour-Nordtvedt-1993}.  Currently, the most precise value for this parameter, $\gamma -1 = (2.1\pm2.3)\times10^{-5}$, was obtained using radio-metric tracking data received from the Cassini spacecraft \cite{Bertotti-Iess-Tortora-2003} during a solar conjunction experiment. This accuracy approaches the region where multiple tensor-scalar gravity models, consistent with the recent cosmological observations \cite{Spergel-etal-2007}, predict a lower bound for the present value of this parameter at the level of $(1-\gamma) \sim 10^{-6}-10^{-7}$ \cite{Damour-Nordtvedt-1993}.  Therefore, improving the measurement of this parameter would provide the crucial information separating modern scalar-tensor theories of gravity from general relativity, probe possible ways for gravity quantization, and test modern theories of cosmological evolution \cite{Turyshev-etal-2007,Turyshev-2008}. 

The reasons above led to a number of specific space experiments dedicated to measurement of the parameter $\gamma$ with a precision better than $10^{-5}$ to $10^{-6}$  \cite{Turyshev-etal-2007,Turyshev-2008}.  Note that SIM will operate at this level of accuracy and, therefore, the Eddington's parameter $\gamma$ will have to be included into the future SIM's astrometric model and the corresponding data analysis.

\subsubsection{Solar Gravity Field as a Deflector}

To model the astrometric data to the nominal measurement accuracy  will
require including the effect of general relativity on the propagation  of light.  In the PPN framework, the parameter $\gamma$ would be part of  this model and could be estimated in global solutions. The astrometric residuals may be tested for any discrepancies with the prescriptions of general relativity. To address this problem in a more detailed way, one will have to use the astrometric model for the instrument including the information about it's position in the solar system, it's attitude orientation in the proper reference frame, the time history of different pointings and their durations, etc.   This information then should be folded into the parameter estimation program that will use a model  based on the expression,  similar to that given by  Eq.~(\ref{eq:defl_chi_diff}). 

To estimate the expected accuracy of the parameter $\gamma$, we use Eq.~(\ref{eq:diff_opd}) to assume that the single astrometric measurement may be  able to determine this parameter with accuracy: 
{}
\begin{equation}
\Delta\gamma   = \Delta  \theta_0  ~
\frac{r^{\tt SIM}_\odot}{\mu_\odot}~
\frac{\sin\frac{1}{2}\chi_{1\odot}\,\sin\frac{1}{2}\chi_{2\odot}}
{\sin\frac{1}{2}(\chi_{2\odot}-\chi_{1\odot})},
\label{eq:gamma_sun}
\end{equation}
\noindent where $\Delta\theta_0$ is the astrometric error of the measurement.

The relativity test will be enhanced by scheduling measurements  of stars as close to the Sun  as possible. Although SIM will never be able observe closer to the Sun than $45^\circ$,  it will allow for an  accurate determination of this PPN parameter. For the accuracy of $\Delta \theta_0=1~\mu$as at the rim of the solar avoidance angle of $\chi_{1\odot}=45^\circ$,  one  could  determine  this parameter with an accuracy  $\sigma_\gamma\sim   7.19\times 10^{-4}$ in a single measurement. Assuming Gaussian error distribution, the accuracy of this experiment will improve as $1/\sqrt{N}$, where $N$ is the number of independent observations. Therefore, by performing differential astrometric measurements with an accuracy of $\Delta \theta_0=1~\mu$as over the instrument's FoR=15$^\circ$, at the end of the mission (with $N \sim 10000$ observations) SIM may reach the accuracy of $\sigma_\gamma \sim  7.2\times 10^{-6}$  in astrometric test of general relativity in the solar gravity field. SIM will provide this precision as a by-product of its astrometric program, thus allowing for a factor of 3 improvement of the currently best Cassini's 2003 result \cite{Bertotti-Iess-Tortora-2003}.  Such a measurement improves the accuracy of the search for cosmologically relevant scalar-tensor theories of gravity by looking for a remnant scalar field in today's solar system.  

\subsubsection{GR Test in the Jovian and Earth' Gravity Fields}

One could also perform a relativity experiment with Jupiter and the Earth. In fact, for the proposed SIM's observing mode, the accuracy of determining of the parameter $\gamma$ may be  even better than that achievable with the Sun. Indeed, with the same assumptions as above, one may achieve a single measurement the accuracy of $|\gamma-1|\sim 4.0 \times 10^{-4}$ determined via deflection of light by Jupiter. As the astrometric observations in the  Jupiter's vicinity will  require careful planning thereby minimizing the number of possible independent observations.  As a result, the PPN parameter $\gamma$ may be obtained with accuracy  of about $\sigma_\gamma\sim 1.3 \times 10^{-5}$ with astrometric experiments in Jupiter's gravity field (note that only $N \sim $ 1000 needed).   
 
Lastly, let us mention that the  experiments conducted in the Earth's gravity field,  could also determine this parameter to an accuracy 
$|\gamma-1|\sim8.9\times 10^{-3}$ in a single measurement (which in
return extends  the measurement of the gravitational bending of light
to a different mass and distances scale, as shown by  \cite{Gould93}). One may expect a large statistics gained from both the astrometric observations and the telecommunications with the spacecraft. This, in return,  will significantly enhance the overall solution for $\gamma$  obtained in the Earth' gravitational environment.

\subsection{Solar Acceleration Towards the Galactic Center}

The Sun's absolute velocity with respect to a cosmological
reference frame was measured photometrically: it shown up as the dipole anisotropy of the cosmic microwave background \cite{Spergel-etal-2007}. The Sun's absolute acceleration with respect to galactic frame can be measured astrometrically: it will show up as a dipole vector harmonic in the global pattern of proper motion of quasars. 

The aberration due to the solar system's galactocentric motion will not be observable because its main contribution is static. However, the rate of this aberration will produce an apparent proper motion for the observed sources \cite{Kopeikin-Makarov-2006}. Indeed, the solar system's orbital velocity around the galactic center causes an aberrational affect of the order of 2.5 ~arcmin. All measured star and quasar positions are shifted towards the point on the sky having galactic coordinates  $l=90^\circ, ~b=0^\circ$. For an arbitrary point on the sky the size of the effect is $2.5 ~\sin\eta$ ~arcmin, where $\eta$  is the angular distance to the point $l=90^\circ, ~b=0^\circ$. The acceleration of the solar system towards the galactic center causes this aberrational effect to change slowly. This leads to a slow change of the apparent position of distant celestial objects, i.e. to an apparent proper motion.

Let us assume a solar velocity of $220 $~km/sec and a distance of 8.5 ~kpc to the galactic center. The orbital period of the Sun is then 250 million years, and the galactocentric acceleration takes a value of about $1.75 \times 10^{-13} $~km/sec$^2$. Expressed in a more useful units it is 5.5 ~mm/s/yr. A change in velocity by 5.5 mm/sec causes a change in aberration of the order of 4 $\mu$as. The apparent proper motion of a celestial object caused by this effect always points towards the direction of the galactic center. Its size is   $~4 \sin\eta ~\mu$as/yr, where $\eta$ is now the angular distance between the object and the galactic center.
 
The above holds in principle for quasars, for which it can be assumed that the intrinsic proper motions (i.e. those caused by real transverse motions) are negligible. A proper motion of   $4 ~\mu$as/yr corresponds to a transverse velocity of $2\times 10^4$~ km/sec at $z=0.3$ for $H_0$=100 ~km/sec/Mpc, and to $4\times 10^4 $~km/sec for $H_0=50 $~km/sec/Mpc. Thus, all quasars will exhibit a distance-independent steering motion towards the galactic center. Within the Galaxy, on the other hand, the effect is drowned in the local kinematics: at 10 ~pc it corresponds to only 200 m/sec.

However, for a differential astrometry with SIM this effect will have to be scaled down to account for the size of the field of 
regard \cite{Turyshev98}, namely $2\sin\frac{\tt FoR}{2}=2\sin \frac{\pi}{24}=0.261$. This fact is reducing the total effect of the galactocentric acceleration to only  $~\sim 1 \sin\eta
~\mu$as/yr ~and, thus, it makes   the detection of the solar system's galactocentric acceleration  with SIM  to be a quite problematic issue.
 
\section*{Discussion}

General relativistic deflection of light  produces  a significant contribution to the future astrometric observations with accuracy of about a few $\mu$as.  In this paper we addressed the problem of light propagation on the gravitational field of the solar system. It was shown that for high accuracy observations it is necessary to correct for the post-Newtonian deflection of  light by  the monopole  components of gravitational fields of a large number of celestial bodies in the solar system, namely the Sun and the nine planets, together with the planetary satellites and the  largest asteroids (important only if observations are conducted in their close proximity). The most  important fact is that the gravitational presence of the Sun,  Jupiter and the 
Earth  should  be always taken into  account, independently on the positions of these bodies relative to the interferometer. It is worth noting that the post-post Newtonian effects due to the solar gravity are unlikely to be accessible with  SIM. This effect as well as the effect of gravitational deflection of light caused by the  mass quadrupole term of the Sun are negligible at the  level of expected  accuracy. However, deflection of light by some planetary quadrupoles may have a big impact on the astrometric accuracy. Thus, the higher gravitational multipoles should be taken into account when observations are conducted in the close proximity of two bodies
of the solar system, notably Jupiter and Saturn.

We emphasized the need of development of a general relativistic model for SIM observables to enable the mission to improve the current astrometric accuracy by a factor of over 1000. This model would have to account for a number of dynamical effects both external to the spacecraft (e.g., motion with respect to the solar system barycentric reference frame, effects of time-varying gravitational field in the solar system (due to planetary motion nd rotation)  on light propagation, various interplanetary media effects, etc.) and internal to the spacecraft (e.g., systematic effects introduced by the spacecraft itself).  Some of this work has already begun in the context of the development relativistic reference frames for the need of future high-precision observations \cite{Kopeikin-Makarov-2007}. However, a lot more efforts is needed; this paper intends to motivate initiation of such a work in the near future.  

\section*{Acknowledgments}
The reported research   has been done at the Jet Propulsion
Laboratory,  California Institute of Technology, which is under  contract to the 
National Aeronautic and Space Administration.


\end{document}